%% file: paper.tex
\documentclass{sig-alternate}
\usepackage{algorithm, algorithmic,rotating}
\usepackage{url,  xspace, cite, shortcuts, subfigure}
\usepackage{multirow}

\usepackage{enumerate}
\usepackage{times}

\usepackage{hyperref}

\pdfpagewidth=\paperwidth
\pdfpageheight=\paperheight

\usepackage{pifont}

\newcommand\comment[1]{}

\begin{document}

\title{Energy Attacks on Mobile Devices}


%
%
%
%
%

\numberofauthors{3} 

{
}

\author{
\alignauthor
Ashish Kundu\\
       \affaddr{IBM T J Watson Research}\\
       \affaddr{Yorktown Heights, NY}\\
       \email{akundu@us.ibm.com}
\alignauthor
Zhiqiang Lin\\
       \affaddr{University Texas at Dallas}\\
       \affaddr{Dallas, USA}\\
       \email{zhiqiang.lin@utdallas.edu}
\alignauthor Joshua Hammond\titlenote{The author contributed to this work and paper during the first half of 2013 when he was a student at Univ. of Texas at Dallas.}\\
       \affaddr{University Texas at Dallas}\\
       \affaddr{Dallas, USA}\\
}

\maketitle
\begin{abstract}
\input{abs.tex}
\end{abstract}

\section{Introduction}
\label{sec:intro}
\input{intro.tex}


\section{Taxonomy of Energy Attacks} \label{sec:taxonomy}
\input{taxonomy}

\section{Building an Exploit} \label{sec:building}
\input{building}

\section{Elements of Energy Attacks} \label{sec:elements}
\input{elements}

\section{Evaluation}
\label{sec:evaluation}
\input{evaluation.tex}


\section{Related Work}
\label{sec:related}
\input{related.tex}

\section{Conclusions}
\label{sec:conclusion}
\input{conclusion.tex}

\bibliographystyle{abbrv}
\footnotesize{
\bibliography{paper,new}  
}

%
%


\end{document}

%% file: abs.tex
All mobile devices are energy-constrained. They use batteries 
that allows using the device for a limited amount of time. In general, energy
attacks on mobile devices are denial of service (DoS) type of attacks. While
previous studies have analyzed the energy attacks in servers, no existing
work has analyzed the energy attacks on mobile devices. As
such, in this paper, we present the first systematic study on how to
exploit the energy attacks on smartphones. In particular, we explore
energy attacks from the following aspect: hardware components,
software resources, and network communications through the design
and implementation of concrete malicious apps, and malicious
web pages. We quantitatively show how quickly we can drain the
battery through each individual attack, as well as their combinations.
Finally, we believe energy exploit will be a practical attack
vector and mobile users should be aware of this type of attacks.

%% file: intro.tex
Smart mobile devices are everywhere: iPhones, Android phones, Windows phones, Blackberry, iPads, Android
tablets, Smart watches, Google glasses, so on and so forth. The computing power of these devices is on a
steep increase day-by-day with multi-core devices already on the market such as
iPhone 4S 
and iPhone 5. 
The power of these devices and
what they offer is being enabled by millions of apps, readily available and inexpensive for the majority of users. With such a thriving ecosystem, the security threats that come with these devices and
apps are also on a steep rise. Android app stores have been found to contain
mal-apps, re-packaged apps that include malwares~\cite{zhou2012dissecting}.
Recently, Apple's messaging system iMessage for their iOS devices were compromised and DoS attacks on some
devices using iMessage were
initiated~\cite{imessage}. It may not be imprudent to say that ``security issues are abundant in an ecosystem that enables freedom
of sharing, and growth for its producers and consumers''. There is a plethora of
works that have studied security of smart mobile
devices (e.g., ~\cite{zhou2012dissecting}).
 
 A common aspect of all such devices is they are energy-constrained, unlike
 their desktop and server counterparts. In particular, these devices include a battery that
 powers the computation and all its usage. If the battery is drained completely
 or to an unusable level, neither the user nor the apps installed on it can use
 the device in the way(s) they need to. What if an attacker manages to carry out
 attacks that drain the energy of a device so that the device is impaired and cannot be used
 until it is recharged again? In this paper, we focus on this type of
 attacks, which we refer to as ``energy attacks on smartphones.'' There have
 been previous works related to energy attacks on server systems by Wu et
 al.~\cite{wu2011energy}. Also, there have been works focusing on energy estimation of
 Android usage~\cite{pathak2012energy,Paul2010androidenergy}. However, existing works have not
 focused on how attackers can carry out energy attacks. 
 
 Consider the following use-case of carrying out such an attack -- an espionage scenario (relevant to a Bond or a Bourne Identity movie) -- ``To kill a device'': a spy is
 waiting for a directive from her boss on her smartphone. The directive can arrive in any way --
 as a call, a message and at any point in next one hour of train ride. The mafia,
 who want to prevent the spy from carrying out the directive, have to succeed. One way
 to enable that is to try to kill the spy, a high risk-high return proposition.
 The other way to enable that is to ``kill the device'' via energy attacks. The
 mafia uses its pre-created hundreds of accounts on google voice, and makes
 calls to this device continuously from these google voice accounts. The spy
 cannot shutdown her device, or disable calling/messaging capabilities because
 she has to receive the directive via the smartphone. The continuous calls lead
 to ``vibration'' and/or ringing of the device, leading to energy drainage, and
 eventually ``killing the device'' by draining most of its energy. 
 
 Such an attack can be easily thwarted by enabling a white list of who can call
 and who cannot call. However, if an  app used by the spy has such a
 vulnerability or has been developed with a backdoor to carrying out energy
 attacks -- such an access control would not work. The question is how effective
 such app-based attacks would be? This paper delves deeper and studies the
 different components of Android operating system, its energy usage and how fast
battery can be drained on Android devices. 
Android commands the smartphone industry as the most-used operating system, as
of now. An attacker could
potentially include malicious code into an otherwise benign application in an
attempt to deny access to a user. Even advanced users may be unaware of how much
battery power each component can drain, and how quickly could a malicious
application could drain the battery. Through our experiments and analysis we seek to
understand the elements of a successful energy attack.\\

\paragraph{Our contributions}: We present a taxonomy of the energy attacks on
mobile devices (\S\ref{sec:taxonomy}), and use it to guide the development of
energy exploits (\S\ref{sec:building}) by systematically exploring hardware,	
software, and network elements (\S\ref{sec:elements}). Our experimental result
(\S\ref{sec:evaluation} shows that our energy attacks are practical, and can
drain battery in a few minutes with a combination of multiple exploits.

%% file: taxonomy.tex
In this section, we first present the principles behind energy attacks in
\S\ref{sub:principles}, then describe why energy attacks are useful with several
compelling use cases in \S\ref{sub:motivations} and the delivery modesl for energy attacks in \S\ref{sub:delivery}, and finally present the taxonomy of such attacks including the elements of on how to carry out energy
attacks and how to construct the exploits in \S\ref{sub:taxonomy}.

\subsection{Principle of Energy Attacks}
\label{sub:principles}

{\em Energy attack is carried out as and when any consumption and/or depletion
of energy on a device is effected by processing of operations with no
legitimate purposes.}

Let $\mathcal{A}$ be the set of operations that processing of operations and $D$ be a device. Suppose
$\alpha$ $\in\ \mathcal{A}$ on $D$ leads to depletion of energy level
by $\delta\ > 0$ amount, then $\alpha$ is a candidate operation for carrying out
energy attacks. An energy-attack comprises of carrying out one or more candidate
operations on the target device with an objective to deplete the energy of a device
by an amount that is greater than zero and less than or equal to $100\%$. If the
event that triggers processing of a subset of candidate operations drawn from
$\mathcal{A}$ on $D$, has its origin in a malicious intention, then the
process of energy depletion on the device is called an energy attack. If the event
does not have its origin in a malicious intention, then it could be called a
bug.

In current computational model, we can safely assume that all operations are
candidate operations including the ``NOP'' operations implemented at the hardware
level. ``NOP'' operations do not lead to any operation being carried out, but it
definitely consumes at least one CPU cycle, and thus leads to consumption of
energy, which would lead to depletion of energy on the device (theoretically
carrying out some number of NOP operations is a form of energy attacks.

A practical energy attack would associate a temporal constraint on the amount of
energy being depleted: the rate of consumption of energy by a specific
combination of the candidate operations is defined as the ratio of amount of energy
consumed over a period of $t_{\delta}$ time and $t_{\delta}$. The greater the rate
of consumption of energy, the more the efficacy of the combination in terms
of energy attack.

\subsection{Malicious Motivation and Use-cases}
\label{sub:motivations}

\input{fig-section2}

The question is why is an ``energy attack on devices'' an appealing attack, and
what would an attacker gain from carrying out such an attack? Let us describe
some other use-cases in addition to the espionage one described in \S~\ref{sec:intro}:

\begin{itemize}
  \item To kill the devices that are parts of a botnet: In order to control the
  extent of  damage a network of device-based bots can carry out, the immediate
  mitigation maybe to shut those devices down or remove them from the network
  (as is done in the case of desktops and servers). However, unlike the desktops
  and servers, devices are mobile, move from one network to another, from 4G to
  WiFi and thus it is harder to control them. Therefore one feasible option is
  to ``drain the energy of those devices remotely'' by carrying out energy
  attacks.

  \item If law-enforcement gets to know the devices being used by terrorists posing an
  imminent threat of a terrorist attack, in order to disband their strategy and
  prevent them from communicating with each other, a stealth energy attack could
  be launched on those devices by law-enforcement. 

  \item Mobile devices are being used to detonate bombs. If law-enforcement know the
  mobile devices that are attached to the bomb, or being used to detonate the bombs,
  perhaps a way to prevent the bomb from being detonated is to drain the
  device of energy.

  \item Protection of a personal device after being lost or stolen: it is common
  to lose or get one's device stolen. Devices contain sensitive personal and
  corporate data; many of the smart devices today have active sessions with web
  services such as gmail, facebook; the devices are configured to ask for a
  password at startup. The owner of the device would want to minimize the risk
  of the device being on the hands of a malicious user X. By carrying out an
  energy attack on the device, the owner would accomplish some or all of the
  following: (1) prevent the user X to gain access to the device at all as a
  restart would ask for a password, (2) prevent the user X to gain access to the
  device at least for sometime until it is charged again, (3) prevent the user to
  gain access to the active sessions, which are lost after the device re-starts.

  \item Business competitors carry out stealthy attacks: Consider that one of
  the businesses in the smartphone market wants to show that its battery is
  more powerful than the a  competitor. In order to bolster its claim among users
  of  the competitor's device, it could launch energy attacks remotely with an
  objective to drain the energy quicker than it does for common user activities.
  That would demonstrate the battery life and convince the users that the battery of the
  competitor's device is not up to the expectation.

  \item A futuristic scenario: As vehicles involve computing and smart-networks
  as part of enabling vehicular networks, they would also rely on battery power to remain operable
  when not being driven. For example, in order to prevent someone from
  traveling, one may need to jeopardize their car. If the car is not within
  physical control, can someone attack its battery by sending multiple network
  packets or unlocking requests, or locking requests? Once the battery dies,
  perhaps that would make the vehicle, which is highly reliant on the battery, remain
  nonoperational until the battery is recharged again.
\end{itemize}

\subsection{Delivery Models of Energy Attacks}
\label{sub:delivery}

Attackers can make use of the following delivery and business models: 
\begin{itemize}
  \item Energy-Attacks-as-a-Service: As a cyber-underworld proposition or
  available to common users, such a service would allow attackers to carry out targeted
  energy attacks for  monetary payments or similar other services as
  exchange. For legitimate usage, such a service would require a requesting
  party to verify its identity against the device (that it owns or has
  authoritative power) on the device, or it has a court order to do so, or
  certain government agencies have provided the power to carry out such an
  attack. As a business model, the attackers can use a subscription-based
  business model, or pay-per-use.

  \item Back-doors for efficient energy attacks: Devices can be designed with
  backdoors available to legitimate parties for energy attacks. The legitimate
  parties could be the owner, network providers, vendor or government
  agencies. However, such backdoors need to be enabled with stringent access
  control in order to prevent exploitation by malicious attackers.
\end{itemize}

Some concerns against energy attacks: Energy attacks could be ineffective when
the device is being charged. Moreover, even if there is value that can be gained
out of ``killing one's device'' via energy attacks, how effective would be
carrying out such attacks? How much time would it require for an attacker to drain
the energy of a device with the available exploits? What information is
essential for the attacker to know about the real-time status of the device in
order to make the attack successful?

\subsection{Taxonomy}
\label{sub:taxonomy}

In this section, we present the taxonomy of energy attacks. The following
Figure~\ref{fig:taxonomy} specifies the taxonomy, and the different
technical aspects of the attack -- why and how. Each column describes the attack
in a more detailed manner. 

Energy attacks are classified with respect to the following parameters:
\begin{enumerate}
  \item {\bf Goals}: There could be four different goals depending how much energy is drained. 
The first one is to completely drain
  the energy of the device (kill the device). The second one is to drain the
  energy partially by a certain percentage or by certain rate or by some specific
  amount. The third goal is to start and/or stop draining the energy of a device
  based on some events, and the last one is to drain the energy of the battery so
  that the battery performance is ruined over a period of time.

  \item {\bf Targets}: Targets of the attack can be one specific device or more
  than one devices, belonging to one individual or group or satisfying some
  constraints, such as location. The attack could be un-targeted -- any
  device that satisfies a constraint such as having an app or visiting a webpage
  can be attacked. 

  \item {\bf Control}: ``Is the attack controlled by the attacker or
  uncontrolled'' is an important aspect of the attack. Whether the attack can start or stop
    an attack and select or de-select some targets at the attacker's will
    defines whether it is controlled or not.

  \item {\bf Locations of launch}: Where is the attack being launched from --
  where is the attacker? The attack is definitely occurring on the device,
  but it may be launched from an app or webpage on the device. It maybe launched from other devices in close
  proximity, or remotely over the network. 

  \item {\bf Elements of the attack}: In order to build an exploit, several
  elements need to be used together. Hardware resources such as GPS, sensors and
  their operations could be used. Software resources such as system calls,
  API, memory allocation and de-allocation, locking and unlocking can be used
  for building an attack. Network-level operations such as data transfer,
  control operations such as handshaking protocols and various network resources
  such as the bandwidth and antenna could be used to build energy attacks. 

  \item {\bf Process of the attack}: An exploit is built, but how to actually
  implement the attack on a specific target in a controlled or un-controlled
  manner? Devices could act as attack origins; humans may be involved in
  launching and controlling the attack; automated agents could be involved. The
  attack could be delivered in stealth mode so that the user could not discern
  whether there is an attack going on or it is the common behavior of the
  device battery; what are the attack policies -- how, when and from where to
  deliver the attack, and how to monitor and manage the attack -- based on
  what events? A workflow may be used to describe an attack delivery. For example, an
  attack is delivered by sending a message to the user about a free app, and after
  the user installs it, the app then talks to its command center on an
  attacker device on how to start the attack and how to monitor such an attack.
\end{enumerate}

\paragraph{Degree of Stealthiness} To determine whether an attack is stealthy or not stealthy at all, we have
defined a scale of detectability of energy attacks (by detecting use of some
components):
\begin{enumerate}
\item Detectable from a distance (stealthiness: 0): a component could be
detected to be used if the device were, for example, in the user's pocket or purse. 
\item Detectable during normal usage (stealthiness: 1): includes 
things like a noticeable change in responsiveness of the device. 
\item Detectable with built-in applications (stealthiness: 2): the device comes
standard with some program or monitor that detects high usage of a component. 
\item Detectable with third party applications (stealthiness: 3): a component's
usage is easily detectable with applications available through the standard marketplace. 
\item Detectable with access to low level functions or system access
(stealthiness: 4): to detect a component the user may have to write a program
to read and analyze some measurements, or may have to access system logs or process information in order to find 
which component is using significant battery power.
\end{enumerate}

 
 For the first two levels, we would expect a component's usage to be detected by
 almost every Android user. To detect attacks that are at the third and fourth
 level, the user needs to have certain minimal knowledge of the device -- such as
 which applications could be used to monitor CPU usage and battery drainage. The
 user needs to realize that the battery is being drained more than usual, and
 needs to know where to look for these applications and how to use them. At the
 fifth level we would only expect a user to detect the drain if the user has
 some familiarity with development on the Android platform. At the fifth level,
 a user would likely need to actively monitor battery usage and then need the
 knowledge and experience necessary to write a new application to detect it, or
 to access system resources to determine the cause. 
 
 We decided that this would
 be the toughest level as we do not foresee any component being completely
 unnoticeable or even requiring hardware access to monitor and detect. Because
 each level requires more user involvement and skills on the part of the user
 from 1 to 5 in that order, we use only the lowest level for each component. This will give an idea of the
 very minimum amount of knowledge and skill required to detect the use of a component that has a large
 impact on the battery. For this part of our testing we used our application to
 turn on each component but did not set a certain battery level threshold. While each component is
running we attempt to detect that the particular component is running either at
an abnormal level, or running when there should not be an application using that component ( other than our
 own). We start by attempting to detect it at level one. If significant usage is
 detected we stop and record the level at which the usage was detected. If,
 after a reasonable effort, the usage is not detected, we go up to the next step
 and again attempt to detect the component's usage.  For the purpose of this
 research we do not attempt to find which application is using the component,
 but merely that the component is being used in a way that may drain the
 battery. Finally we give information on each component that may make use of the
 component more difficult or preventable. We specifically make note of settings that may
need to be on for a component to work. This is closely coupled with the
permissions required for an application to make use of each component. Knowing
this makes it much easier to detect an application that may have a large impact
on battery before it is installed. This information allows users to make a more informed decision about whether an
 application can use the access rights, that the users gives to it, to drain the
 battery.

%% file: fig-section2.tex
%

\begin{figure*} [t]
\begin{small}
\centering
\includegraphics[width=6.5in]{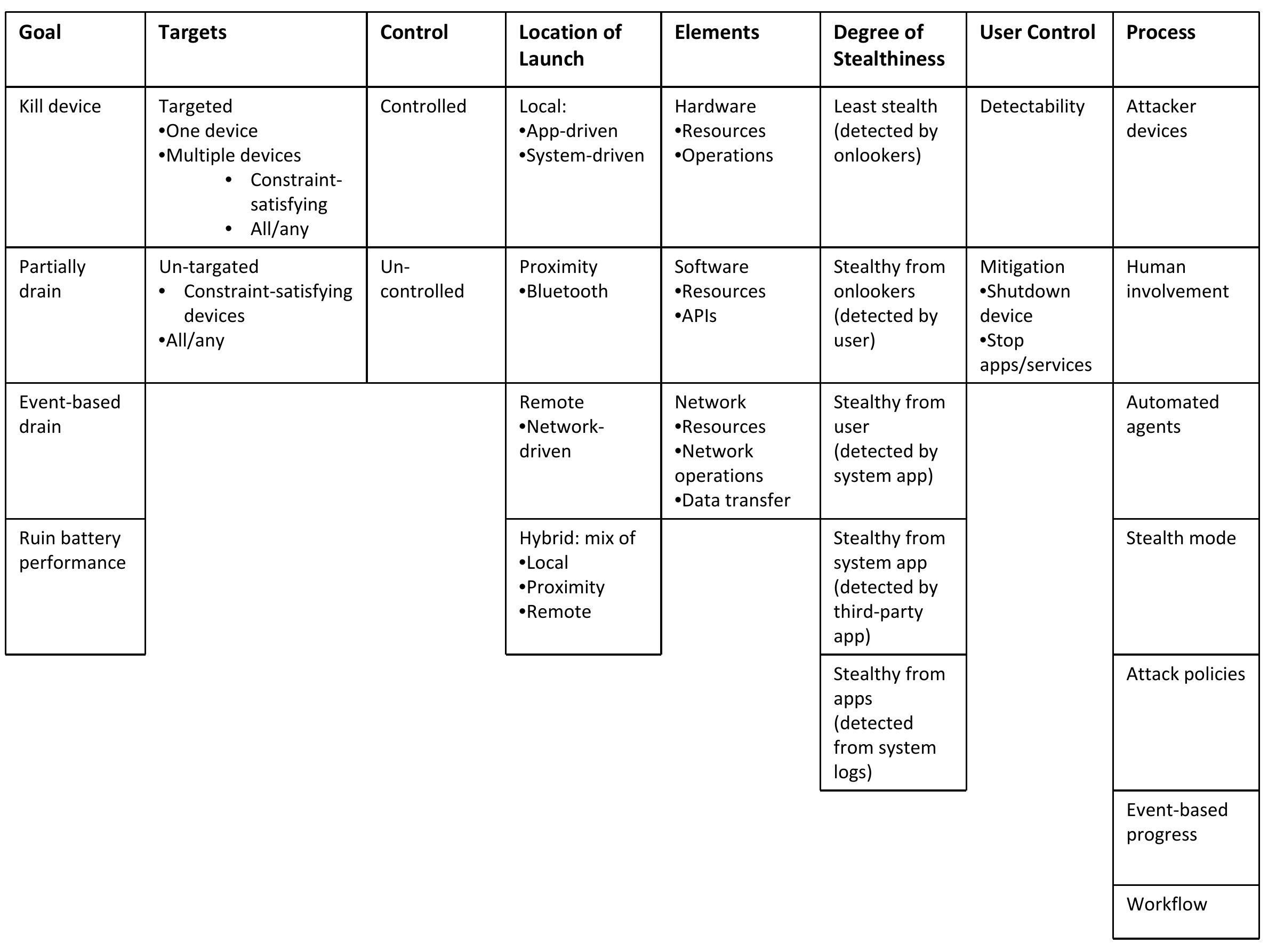}
\caption{Taxonomy of energy attacks}
\label{fig:taxonomy}
\end{small}
\end{figure*}

%
%
%
%
%
%
%
%
%
%
%
%
%

%% file: building.tex
Building an exploit for energy attack does not depend on traditional
vulnerabilties in software stack, hardware devices or in cryptography. That
makes it easy to build such an attack and hard to defend. 

An exploit uses each of the aspects of the taxonomy discussed earlier. An
attacker decides the goal(s), the targets of the attack and
controls on the attack. Based on the available methods of attack delivery, the
attacker can decide the elements to be used for the attack. If the attack
delivery method available to the attacker is via installing an app on the device
via a promotional ad, the attacker then goes ahead and implements the app that
uses several elements of the attack. If the attacker has access to a backdoor on
the device, then based on the elements exposed via the backdoor,  exploit can
be built.

Delivering the attack after building an exploit can be implemented in several
ways. By getting the user to install the app by luring it for a coupon, or by
some other social engineering method, the attack via the app can be delivered.
The delivery model then can be Energy-attack-as-a-Service or via backdoor
exposed via the app. How to get the attack delivered is part of  building the
attack; however, due to constraints on the page length, we would
not go into further details.

\paragraph{Delivering the exploit to the device}
Several techniques can be used to deliver the exploit to the device(s). Some of
the ways to attack a device are:

{\em Attacker knows the phone number:}
\begin{enumerate}
  \item Attacker sends a message with a coupon or similar monetary incentive and
  asks the user to install the app or visit website to get the coupon.
  \item App: delivers more control on the attack. An app could include the exploits during its development, or introduced during the lifecycle.
  \item Visiting the website delivers the attack only for the time of visit.
\end{enumerate}

{\em Attacker knows the user's name:}
\begin{enumerate}
  \item If the attacker is a social network or one of its admins or a gaming
  service that the user has enrolled in, and has its app installed on the
  device, or if the attacker can carry out a network-based man-in-the-middle
  attack by injecting traffic:
  \item the attacker pushes data, scripts and events to trigger the app.
  \item the attacker then can lure the user as in previous example to install
  other apps or visit webpages.
\end{enumerate}

{\em Attacker knows the webpages the user visits:}
\begin{enumerate}
  \item The attacker controls the webpages, then it can add energy-consuming
  software and network elements. In particular, a web-page may contain java-scripts that carry out energy attacks. 
  These java-scripts could contain infinite loops, or recursive code, or data download, malicious applets, etc., to drain the battery.
  \item The attacker otherwise can use Google Ads or such other services for
  displaying advertisements including javascripts that do not look malicious but
  carry out energy attacks.
\end{enumerate}

%

%% file: elements.tex
In this section, we describe the elements of the exploit. There are in general three types of
elements: \emph{hardware}, \emph{software}, and \emph{network}.
Hardware elements are rarely exposed as they are -- software elements are almost
always combined with hardware elements. For example, accessing GPS is allowed
from an app, which thus can include some software-based elements in the exploit.
Network-based elements can be used along with the software elements. A software
app can maliciously download more data than needed thus consuming the energy via
network elements and software elements. In order to determine how effective each \emph{hardware}, \emph{software}, and \emph{network} element is towards carrying out energy attacks, we carried out several experiments by developing the exploits as apps, webpages based each elements or a combination of them.  A high level overview of
these elements is presented in Fig.~\ref{fig:element}. In the following, we present how to construct the concrete attacks by using these elements.

\paragraph{Experiment Setup} In order to determine the efficacy of energy attacks by
using the components, we need to determine how long it takes to achieve the
goal of the attack -- draining energy by using the corresponding components. We carried out
two types of experiments: (1) for draining energy from 100\% to 0\% (kill the
device), and (2) an energy depletion by 5\% (partial draining). We also did a
control test. In each of our tests the screen is wake-locked, which means the
display will not turn off and the phone will not enter sleep mode. This is
necessary so that the operating system does not begin powering down components
and give inaccurate readings as to their battery usage. Thereby we also acquire
the wake-lock for the control test. By using this wake-locked control we can see
how much energy is consumed by each component compared to the wake-lock and that
should give us an accurate comparison of which components consume more energy.
\S\ref{sec:evaluation} will present the detailed experimental results. 

\paragraph{Single vs. Combination}: Exploits could be  based on a single element
or a combination of elements, and can be parallel or sequential. Parallel exploits are expected to be more
effective in carrying out energy attacks  because, the  software, hardware or network elements each thread
or parallel process uses consumes power, as well as there is a cost of
scheduling and bookkeeping of threads/processes at the system level. In the
exploits we have experimented with, if it is a combination of elements, then we
have tried to maximize the battery drainage by making it parallel -- either at
process-level or at the app level.

While most energy attacks would be a complex combination of elements of different
categories -- software, hardware, network and would be used both in sequential
and parallel; in order to quantify how effective such exploits would
be, we need to understand how effective and quantify the energy draining
capabilities of exploits developed based on single elements. 

\subsection{Hardware Elements} \label{sec:hw}
\input{hw.tex}

\subsection{Software Elements} \label{sec:sw}
\input{sw}

\subsection{Network Elements} \label{sec:nw}
\input{nw}



%% file: hw.tex
Using hardware components of a device can drain the battery significantly.
Several hardware components/sensors are part of the hardware packaging of the
device. Some such components covering some of the  common components across
devices running Android and iOS: Other than CPU, display and brightness via
ambient sensor, camera, flash, bluetooth, Wifi, 3G or 4G, GPS, gyroscope,
accelerometer, proximity sensor, and magnetometer.
\begin{figure} [t]
\begin{small}
\centering
\includegraphics[width=3.5in]{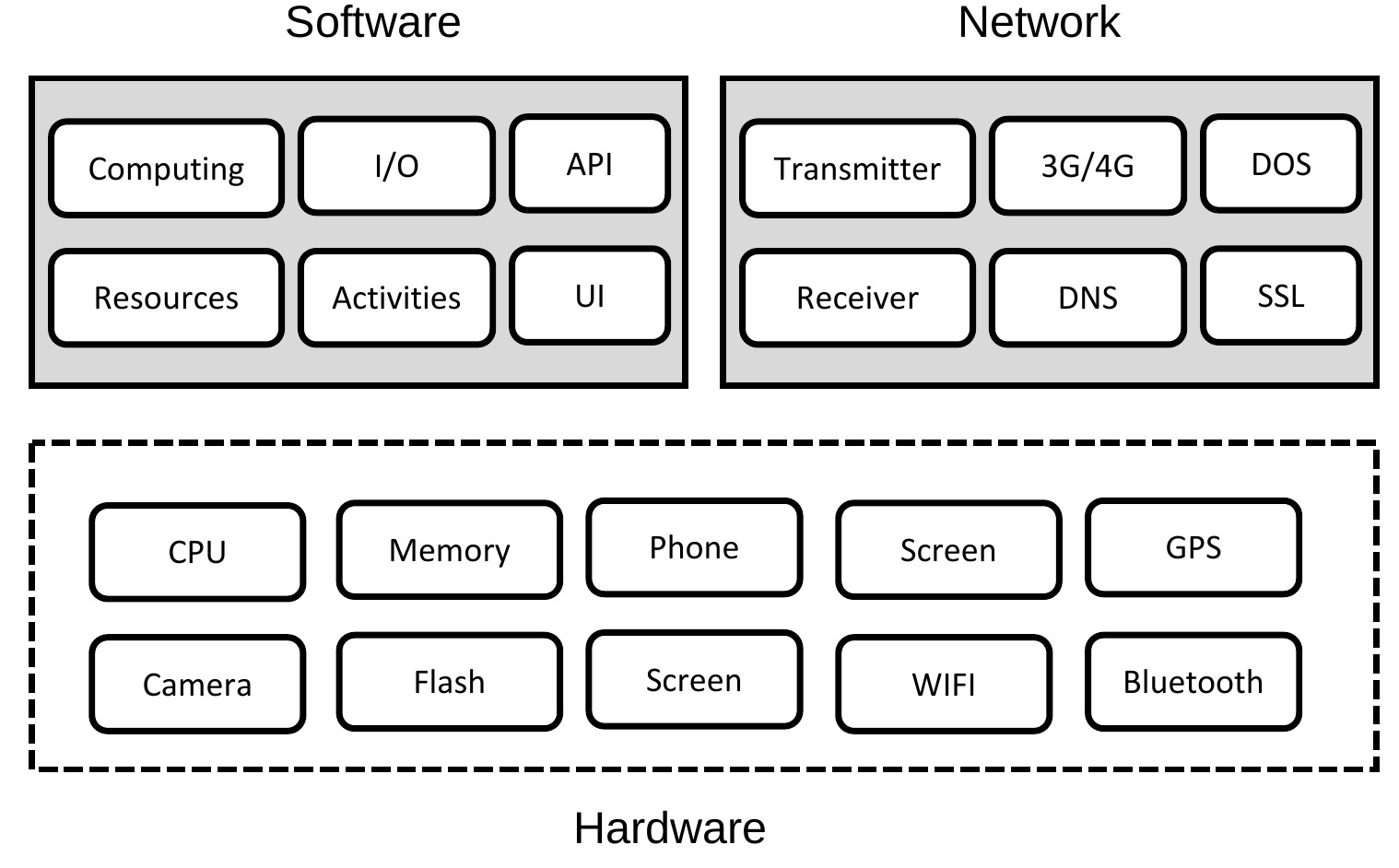}
\caption{Overview of the Exploit Elements}
\label{fig:element}
\end{small}
\end{figure}

The fundamental method of carrying out effective hardware-level energy attack is:
{\em to develop an exploit based on  the
hardware elements that can be used concurrently with minimum authorization
needed, and consume most amount of power when used concurrently together.
Determine these elements  and their combination statically and/or dynamically
depending on the state of the device and the elements.}


\subsubsection{Basic Components}
{\em CPU:} In order to determine how effective  CPU-intensive exploits would be,
 we used an app that carries out matrix multiplication using two threads. The
 matrix multiplication starts with two initial matrices and then continues to
 multiply the resulting matrix by one of the starting matrices. In our
 experiments, this has a CPU utilization of up to 90\%, which is a good measure
 of the efficacy of CPU-intensive exploits. Compute-intensive exploits may or
 may not be stealthy at all depending on how the system handles high CPU
 utilization -- some systems may not respond to user interaction well, which
 makes the attack somewhat less stealthy.
 
{\em Memory}: Memory, especially video memory consumes a large amount of energy.
Playing video would exploit this consumption of energy, which however, would not
be stealthy at all. In our implementation, we loaded a video to the memory, and
played it in an infinite loop. However, playing different videos would be more
effective as it includes writing to the video memory for each new video other
than reading and refreshing the memory.

{\em Phone:} In order to exploit the phone call functionality, an
echo number is used as the other number to be called. The advantages of using an
echo number are: it does not require extra participants to receive a call; it does
not disconnect prematurely. It can be stealthy until the user tries to use the
phone functionality and discovers that a call is being made (phones do
not have the mechanism for multiple active call sessions).

{\em Camera flash:} Camera flash can be used to attack the device, but it is
not stealthy at all. In our exploit, we have an infinite loop that turns
the camera's flash on then off. Theoretically, leaving the
flash light on  will drain less energy than when turned off and on
continuously. 

{\em Camera:} Camera of a device uses several resources and thus an important
candidate for building ane exploit. This element is of medium stealth, but user
interaction with the camera would reveal the app being used.
 To exploit the photo
capability and we used the camera interface, rather than the normal method of
calling another application to take care of the actual photography. This gave us
more control over what components were being used, allowed us to take pictures
rapidly, and allowed us to automate the process.

\subsubsection{Sensors}
There are several other sensors that can be used to build exploits. Android
supports a sensor API framework, which allows the program to determine the power
requirements of each sensor. An exploit may use these APIs and determine the
sensors that require the most power dynamically and start using some of them. In
the following discussion, we describe some of the sensors with which we
have developed exploits.  Other sensors such as accelerometer, gyroscope,
proximity sensors, light sensors, magnetometer, temperature and pressure sensors
can be exploited similarly. Adding a listener, removing a
listener, probing a sensor, acquire the sensor data continuously are different
actions that can be carried out repetitively and in parallel to drain the energy.

{\em Screen brightness sensors:} Screen brightness is highly non-stealthy and
also consumes large amount of energy. For testing screen brightness we turned
the brightness up to its maximum while in our wake locked state. It is important
to note that this test would not be accurate without the wake-lock that keeps
the display active. For this component to drain the battery at the levels we
record in our experiment, the display must remain active throughout the entire
period. It is also important to note that during our experiments we were unable
to turn up the brightness when the battery was low (as controlled by the Android
OS). This also requires that the phone stay awake and may not functionally drain
the battery from the background.

{\em GPS:} GPS component as an element of energy attacks can be effective and
are highly non-stealthy, as most devices show use of GPS on the top-bar of the
phone. To test the GPS component, we enabled a location listener and constantly
read the location of the device. This is similar to how a GPS would have to be
used with a map application, to maintain a current and accurate location. 

{\em Bluetooth}: Switching bluetooth on and off itself can consume energy. This
activity can be a stealthy one as long as bluetooth activity can be hidden from
the display and UI.

%% file: sw.tex
Software elements may be compute-intensive, I/O-intensive,
network-intensive or sensor-intensive, or a mix of both of them. The
software-level elements include OS-level APIs, resources such as locks and
semaphores, threads/processes, libraries, software cache, file system, system
calls,  sockets, events and notifications, and messages. Higher-level elements
are activities, UI components, Java libraries and so on. We have developed
exploits on a combination of some of these APIs. 

In order to develop an effective energy attack based on software-based elements,
a rule of thumb is: 
{\em An effective energy attack uses software elements that can be used
concurrently and that consume most amount of energy when used together.}

Some of the rules of thumbs is to use elements that:
\begin{enumerate}
  \item    lead to as many context switches, memory
  swapping and I/O as possible. 
  \item run stealthily -- in the background and does not
  interfere with useability of the device.
  \item force using system calls and/or lead to interrupts
  \item are complex mathematical operations such as cryptography, matrix
  multiplication, graphics.
  \item belong to different libraries, use data blocks from different memory
  pages thus defeating the locality-of-reference strategy.
  \item start servers that listen to on some ports creating backdoors to the
  device.
\end{enumerate}

In our experiments, we developed exploits based on cryptography,
matrix multiplication, notification, database operations and web-based
operations.

{\em Cryptography}: Cryptographic operations such as encryption,
digital signatures involve complex mathematics and are expensive operations.
They  also run quite stealthily and are known to consume lot of energy
especially when cryptographic accelerators are not in use. Our exploits used
RSA encryption and signature generation.

{\em Database Operations}: Database I/O also is a stealth operation and is
expensive. Table creation, deletion, data addition and removal are to be
included in the operations. If caching can be disabled by the attacker
programmatically with no higher privilege required, it should be disabled.

{\em Notifications}: Notifications are very common in
consumer applications and we thought it would be important to show their impact.
To test the screen rotation we constantly swapped between landscape and portrait
layouts. Though this component was tested through API calls, we believe that it
is also a reasonable estimate for how the device would drain power when the
device is rotated to change screen orientation.  For
our experiment,  we constantly opened and closed a notification.

%% file: nw.tex
The network elements use the transmitters and receivers of the device -- both
for 4G, Wifi and bluetooth capabilities. Network operations also lead to use of
network stack, use of cryptography when SSL is enabled. 

The general principle of development of exploits for energy attacks via network elements
is: 
{\em Use the network operations  such that they can be carried out concurrently
with maximum data transfer such as wifi and bluetooth used concurrently, as much
processing as possible by the network components of the device, synchronize the operations such that the device has
to power on and off its transmitters and receivers as much as possible, and
inject energy attack payloads when the network communication is vulnerable. }

There are several elements related to network and communication that can be
exploited: bluetooth-based communication, wifi and 4g-based communication,
data transfer, control protocols, DNS query attacks and so on.

{\em Bluetooth-based Communication}: Switching bluetooth on and off itself can
consume energy. This activity can be a stealthy one as long as bluetooth activity can be hidden from
the display and UI.

{\em Wifi-download:} For our download test, we have a statically coded
address for an image on the web. This image is constantly downloaded and written
to the SD-card. Though we could have skipped over writing to the SD-card it
seems that a common application would not download something to discard it and
would likely constantly store downloaded data for the application.  

{\em DNS query attack}: DNS queries are sent out by the device in clear, and a
form of Kaminsky attack can be used to drain the energy of the battery
significantly. Moreover, the DNS cache poisoning can be used to make the device
send the data several times, send data to hosts from where the destination is
unreachable. 

{\em DoS Attack}: Carry out DoS attacks on the open ports on the device such as
via SYN-flodding attacks.

%% file: evaluation.tex
In this section, we have described several experiments that we have carried out
to determine effectiveness of several hardware, software and network elements in
carrying out energy attacks. The experiments include exploits developed based on
single elements or a combination of these elements. 

{\bf Experimental Setup}: 
For our experiments we used a Samsung Captivate Glide
(SAMSUNG-SGH-I927) smart phone with Android Gingerbread (Android version 2.3.6).
The hardware capabilities  covered are: CPU, vibration, camera flash, WiFi
download, bluetooth, phone, 4G download, brightness, video playback (video
memory), GPS,  screen rotation, and camera. 

{\em Process of experiments}:
For each test, the app implementing an exploit we developed, allows the tester
to select which components to enable for the test. Then, when the start button is pressed, the
application starts each individual component. After every component has been
started the application records the current time and takes an initial reading of
the battery level (reported in percent). The main application then sleeps for
two second intervals while the component runs. Each time it wakes, it checks the
current battery level. If the difference between the original battery level and
the current level is below the threshold of five percent, then the application
goes back to sleep, with the components still running in the background. If the
threshold has been met then the application records the current time, compares it with the
 starting time and makes a notification to the user about the number of minutes
 the application took to drain to the threshold. For our experiments, we tested
 each component ten times. This allows us to take an average that compensates
 for some of the fluctuations we may see in each individual test.

{\em Most effective exploits:} Top three exploits that drain the battery from
100\% charge to 0\% and that drain 5\% of the battery
 with lowest time required  are described in the following table. The web attack
 that includes an exploit delivered by a webpage took two hours and forty five
 minutes and is the second most effective attack, which moreover can be carried
 out in a stealth manner. The other two attacks are not stealth but the most
 effective attack can drain the battery within one hour and forty five
 minutes.\\ 
 
\begin{tabular}{ |l|l| }
\hline
{\bf Top 3 exploits} & {\bf Time in mins} \\ 
& {\bf for 100\% draining} \\ \hline
1. Brightness, CPU and Camera Flash & 104 \\ \hline
2. Web attack &  164 \\ \hline
3. Screen Brightness & 204 \\ \hline \hline
{\bf Top 3 exploits} & {\bf Time in mins} \\ 
& {\bf for 5\% draining} \\ \hline
1. Brightness, CPU and Camera Flash & 4.8 \\ \hline
2. Web attack & 6.5\\ \hline
3. Brightness, CPU and Camera Flash &  7.2 \\ 
& while charging \\ \hline 
\end{tabular}\\ 

In the following table, we have presented the results of our experiments on the
effectives of exploits. For each component we give five statistics: Average teim
to drain 5 percent, standard deviation of draining time, maximum time to drain 5
percent, and minimum time.  Each
measurement of time is recorded in minutes.\\ 


\begin{tabular}{ | l | l | l | l | l | }
\hline
Component & Average & St. Dev. & Max. & Min.  \\ \hline
Vibration & 19.4 & 1.075 & 21 & 18  \\ \hline
CPU & 9.5 & 0.972 & 11 & 8  \\ \hline
Camera Flash & 9.3 & 1.059 & 12 & 8  \\ \hline
WiFi Down. & 23.5 & 3.598 & 29 & 16  \\ \hline
Bluetooth & 25.2 & 5.514 & 36 & 18  \\ \hline
Phone & 13.8 & 1.932 & 18 & 12 \\ \hline
4G Down. & 11.1 & 1.197 & 13 & 9  \\ \hline
Brightness & 7.4 & 1.075 & 10 & 6 \\ \hline
Video & 16.8 & 1.989 & 22 & 15  \\ \hline
GPS & 17.4 & 1.734 & 19 & 15 \\ \hline
Notification & 26.6 & 4.351 & 33 & 20 \\ \hline
Rotation & 17 & 3.197 & 23 & 13  \\ \hline
Photo & 12 & 1.764 & 14 & 9  \\ \hline
Encryption & 12.3 & 1.059 & 14 & 11 \\ \hline
\end{tabular} \\

\subsection{Hardware elements}
Our results show that the screen brightness drains battery faster than any other component that we tested. This is closely followed by the camera flash 
and CPU usage. Using only the brightness would drain the battery in under two and a half hours. 
For our case of normal usage, this is a rather short amount of time. A smart phone is reasonably 
expected to last at least 16 hours per day (the time a person would be awake with 8 hours of sleep each day).
 This suggests that being able to detect and monitor these components could be very important for normal users. 
 For the case of an attack this may be longer than an attacker would like. If the attacker was attempting to close 
 off usage of the phone for a specific time or purpose, this may be ample time to detect the usage, or at least 
 notice the drain and begin charging the phone.  We also tested the exploit
 while the phone was being charged, and we found that energy attacks indeed can
 succeed while the phone is being charged.

\input{figures}


 The full drain using screen brightness attack took just over 204 minutes
 (Figure~\ref{fig:Brightness}).
In figure ~\ref{fig:DeltaBrightness}, we see a large spike at the end of this
graph for the last few percent of the battery life which helps account for the
longer than expected drain time. This is where Android automatically lowers the
screen brightness to preserve battery life. At that point we have already
drained most of the battery and any extended usage would not be possible. This
drain may take much longer than an attacker would like and suggests that an
attacker would likely need to use multiple components.
 
 To get an idea of how multiple components perform together we tested the three
 top components -- brightness, camera flash, and CPU at once. If they were
 completely independent we would expect that together they would drain five
 percent of the battery in under three minutes. Our test showed that the three
 components drained the battery in, on average, 4.8 minutes. This implies that
 there is some overlapping usage within groups of components. We also did a
 larger scale test of these three components. We started with a full battery and
 ran all three components. For this test we set a threshold of 2 percent, at
 which point the application wrote the battery level and total time to a file.
How this attack drained the battery over time is shown by Figure
~\ref{fig:threeattack}). The battery starts at 100 percent charge and drains
down to 0 percent in just over 100 minutes by this attack. We also see that the
attack drains almost linearly over time. This implies that the reduction in
usage of some of the components at low power does not make a large impact on the
overall attack. However, this did cause the device to become warm to the touch.
Depending on the context of usage of the device, this may be difficult to detect
or could be a major indicator of high battery usage. 

We also
did some tests with our normal 5 percent threshold while the device was
connected and charging via USB. Despite the phone being charged while the attack
was going on, the attack was still able to drain 5\% of energy in an average of
7.2 minutes. This suggests that an energy attack is certainly able to kill the
device even when it is charging. Moreover, Android allows any program to monitor
the battery level and charging state without any extra permissions. 
An attacker could use this as a cue to
 scale up the energy attack when the device is charging to ensure that the
 device is killed.

Attack by taking photos drains the battery from 100\% charge to 0\% in about 265
minutes (Figure~\ref{fig:camera}). By just (stealthily) taking photos using the
camera, an Android phone can be drained out of battery in over four and half
hours (less than the flight time from New York to San Francisco).

\subsection{Software elements}
We carried out experiments demonstrating how attacks using encryption can be
used to drain battery. In Figure~\ref{fig:encryption}, a 100\% charged battery
was drained fully  about 200 minutes. However, the battery was about 95\%
drained at 175 minutes, and during the remaining 25 minutes it drained 5\%
because Android reduced the brightness of the display due to its in-built
policies.  

In another exploit that implemented energy drain via repetitive
database procedure: generation of strings of about 1KB, its addition to a table
in the database and then deletion of the data, the battery was fully drained
within 260 minutes (there is a similar spike towards the end due to savings of
energy by dimming the display by Android) (Figure~\ref{fig:dbdata}). In
Figure~\ref{fig:dbtable}, the battery is drained fully within 300 minutes by
addition and deletion of tables in a database on the device. In order to make
the database-related exploit more effective, we included encryption in the
exploit: data is generated, encrypted, added to the table, and then deleted.
Figure~\ref{fig:dbencr} shows that encryption improved the exploit effectivenes
by about considerable 20 minutes -- it took about 240 minutes. We could add more
cryptography operations to improve this period. 

We also demonstrated how a web-app with a malicious javascript can carry out an
attack. Figure~\ref{fig:web} shows that such an attack is highly effective -- it
took only 164 minutes to drain the battery of the device fully. The javascript
carries out network traffic and CPU usage. The attack
showed no slow down at low power and stayed fairly consistent throughout the
test.   Every time the web application is killed it is brought back up, to the same web page, by our malicious application. 

\subsection{Network Elements}
As described in the table above, Wifi-based data transfer drains 5\% of energy
at an average rate of 23.5 minutes. Bluetooth-based data transfer drains 5\%
energy at an average rate of 25.2 minutes. However, 4G-based data transfer
consumes the most energy at an average of 11.1 minutes over 5\% drain. Due to
space constraints, we could not add the plots related to the  energy attacks due
to network-based data transfer.

{\bf Most efficient attack}: An exploit based on several hardware, software and
network elements is developed. As Figure~\ref{fig:finalattack} shows this attack
takes least amount of time -- 92 minutes to drain the battery from 100\% to 0\%. 

\subsection{Configuration of permissions}
Permissions are not needed for webpages to access default elements. However, for
apps,  permissions  for each hardware element has to be allowed, so that a malicious energy attack app has to choose from those set of components
 it has the privilege of accessing. Generally the setting will have to be
turned on by the user unless the application is given the required permission to
change settings. The following table specifies the permissions required for each
element. The final column marks whether the permission is required even if the
required settings are enabled.

\begin{tabular}{ | l | p{2cm} | p{2cm} |  p{1cm} |}
\hline
Component & Setting & Permission & Req. \\ \hline
Vibration & -- & VIBRATE & Y \\ \hline
CPU & -- & -- & N \\ \hline
Camera Flash & -- & FLASHLIGHT & Y \\ \hline
WiFi Down. & WiFi enabled & CHANGE WIFI STATE & N \\ \hline
Bluetooth & Bluetooth enabled & BLUETOOTH & Y\\ \hline
Phone & -- & CALL PHONE & Y\\ \hline
4G Down. & Mobile data enabled & CHANGE NETWORK STATE & N\\ \hline
Brightness & -- & -- & N\\ \hline
Video & -- & -- & N\\ \hline
GPS & GPS enabled & ACCESS FINE LOCATION & Y\\ \hline
Notification & -- & -- & N\\ \hline
Rotation & -- & -- & N\\ \hline
Photo & -- & CAMERA & Y \\ \hline
Encryption & -- & -- & N \\ \hline
\end{tabular} \\

Entries with '--' indicate that no setting or permission is needed. 
All of the permissions are under android.permission.
All of the components with no required settings or permissions are available to any program under the Android API.
For the case of normal use these permissions should give a good idea of how a desired application may impact battery usage. 
This information can be used to make good decisions about whether an application is a potential risk to battery life. These should also 
inform users on what they may need to monitor in each application to see if it is using excessive battery. \\


battery and their possibility of draining battery without the user knowing. Overall for normal usage the take-away is that it is easy and important to monitor component usage, and have an idea how an application may impact battery life. For an attack it is important to be prepared. Turning off any components that are not being used can help prevent usage by applications without specific permissions. Monitoring programs will help detect abnormal usage if an attack occurs. Disabling the camera, through administrative access, may be an important part of preventing an attack. It is important to note that these can help mitigate the risk but will not prevent it. An application still has access to some resources without any permissions and many components may be hard to detect if the user does not know what they are looking for or does not recognize the change. \\

%% file: figures.tex
\begin{figure*}
\centering
\begin{tabular}{ccc}

\begin{minipage}[b]{0.33\textwidth}
\hspace{-0.16in}
 \subfigure[Energy drain of screen brightness attack over full battery life]{
\label{fig:Brightness}
\includegraphics[width=2.2in]{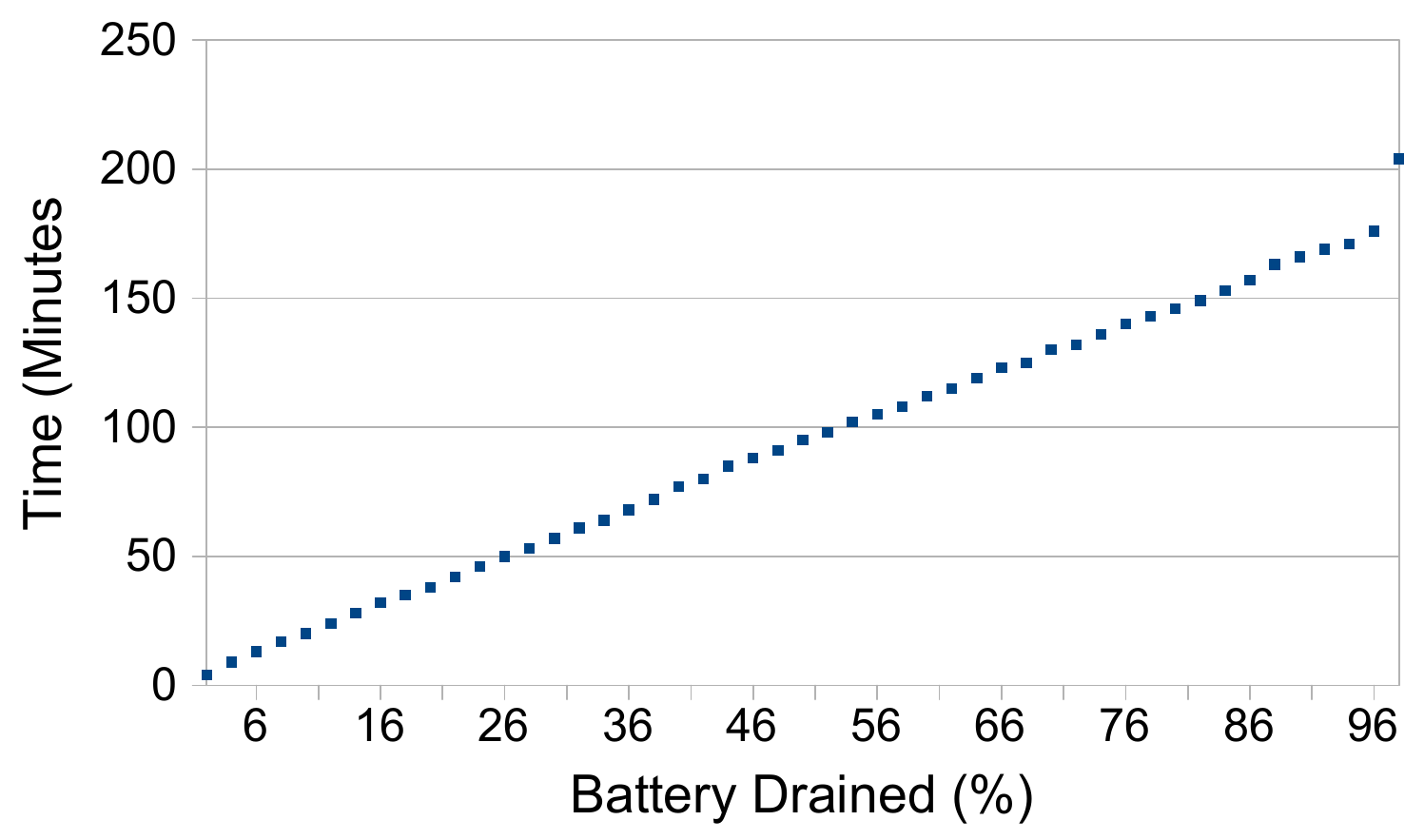}}
\end{minipage}
&

\begin{minipage}[b]{0.33\textwidth}
 \subfigure[Energy drain of 3 component (brightness, flash and
 CPU) attack over full battery life]{
\label{fig:threeattack}
\includegraphics[width=2.2in]{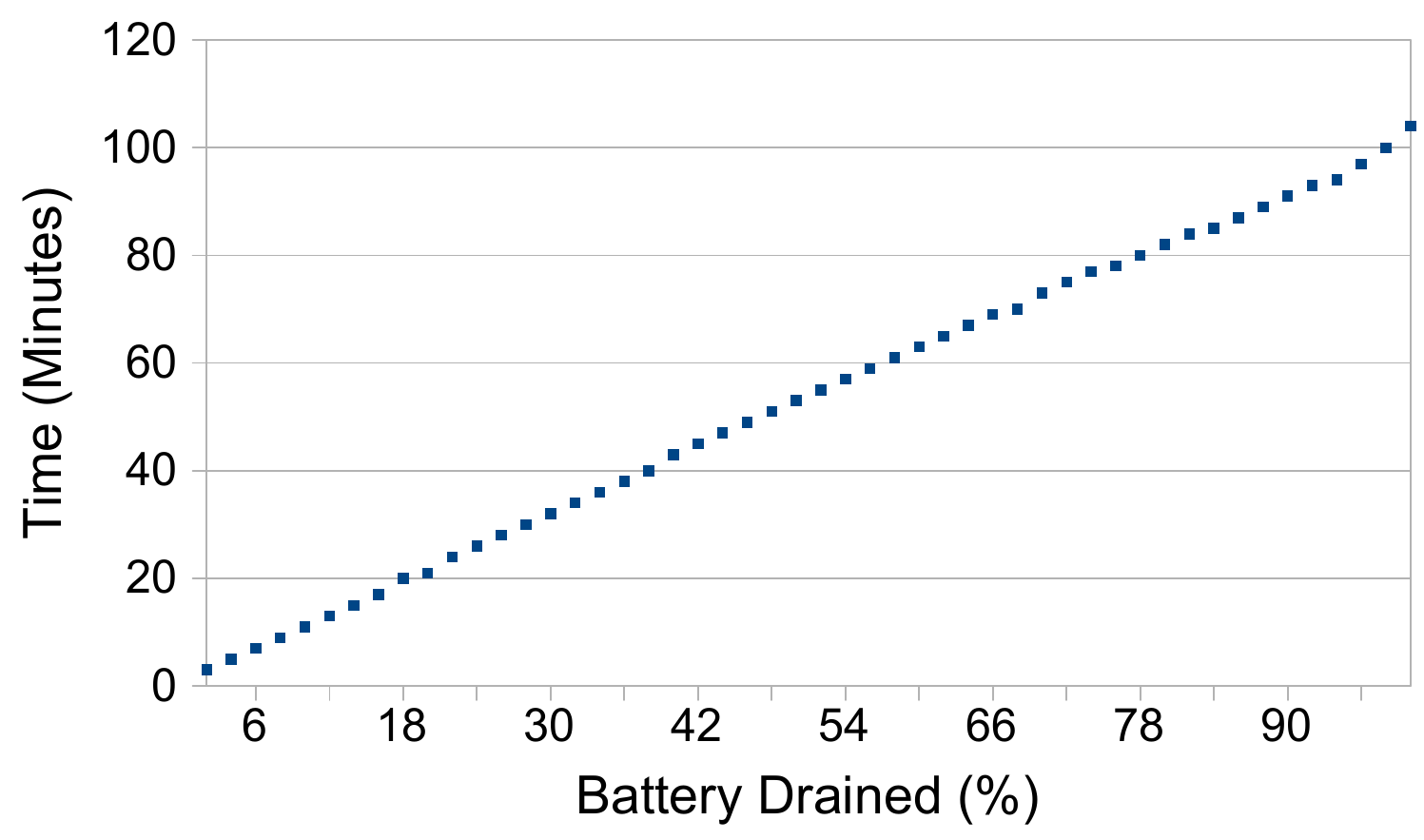}}
\end{minipage}

&
\begin{minipage}[b]{0.33\textwidth}
 \subfigure[Energy drain of camera attack over full battery life]{
\label{fig:camera}
\includegraphics[width=2.2in]{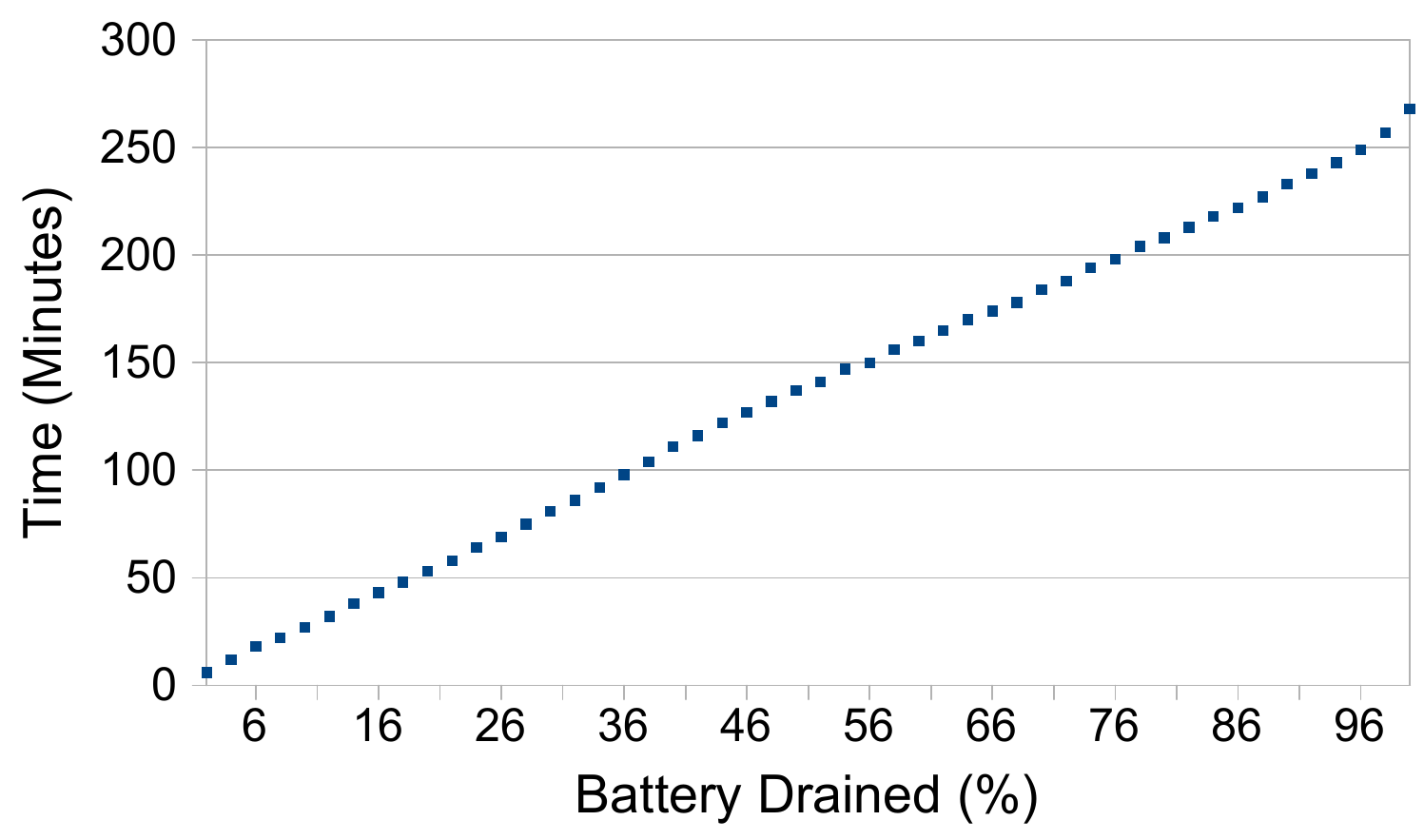}}
\end{minipage}

\end{tabular}
\label{fig:taskprofile} \caption{Exploitation using hardware elements.}
\end{figure*}


\begin{figure*}
\centering
\begin{tabular}{ccc}

\begin{minipage}[b]{0.33\textwidth}
\hspace{-0.16in}
 \subfigure[Energy drain of encryption attack over full battery life]{
\label{fig:encryption}
\includegraphics[width=2.2in]{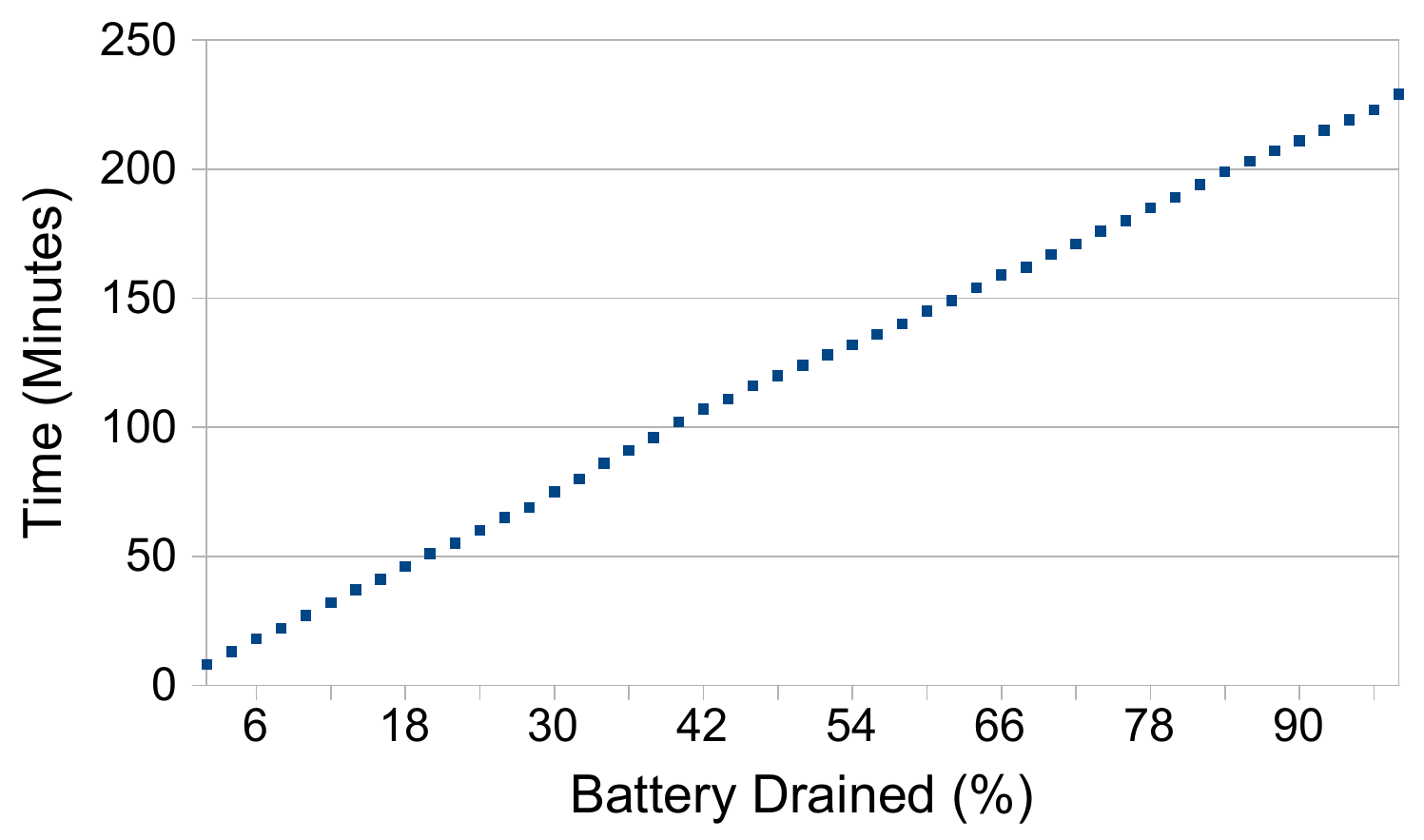}}
\end{minipage}
&
\begin{minipage}[b]{0.33\textwidth}
 \subfigure[Energy drain of database data generation, addition and deletion
 attack over full battery life]{
\label{fig:dbdata}
\includegraphics[width=2.2in]{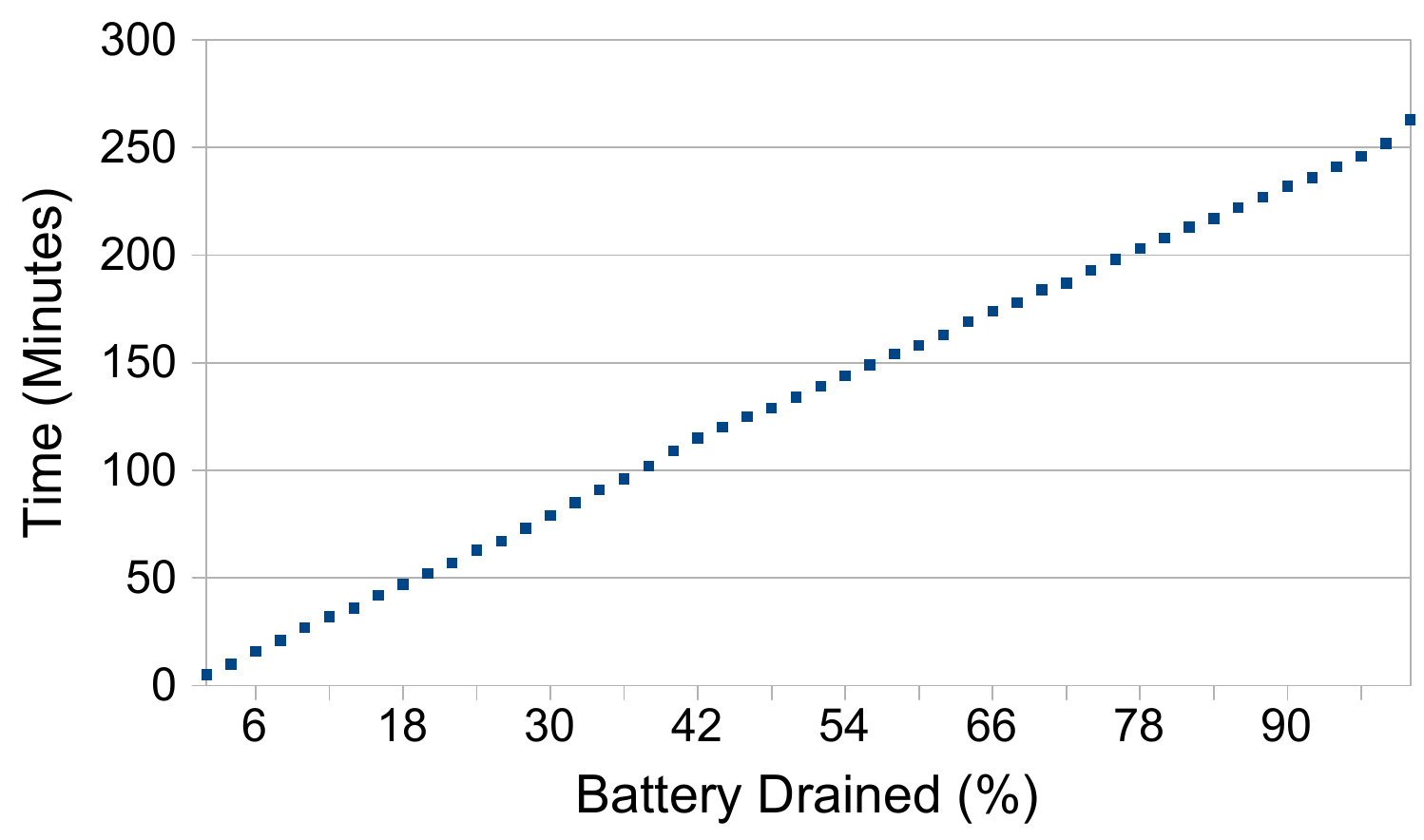}}
\end{minipage}
&
\begin{minipage}[b]{0.33\textwidth}
 \subfigure[Energy drain of database table addition and deletion attack over
 full battery life]{
\label{fig:dbtable}
\includegraphics[width=2.2in]{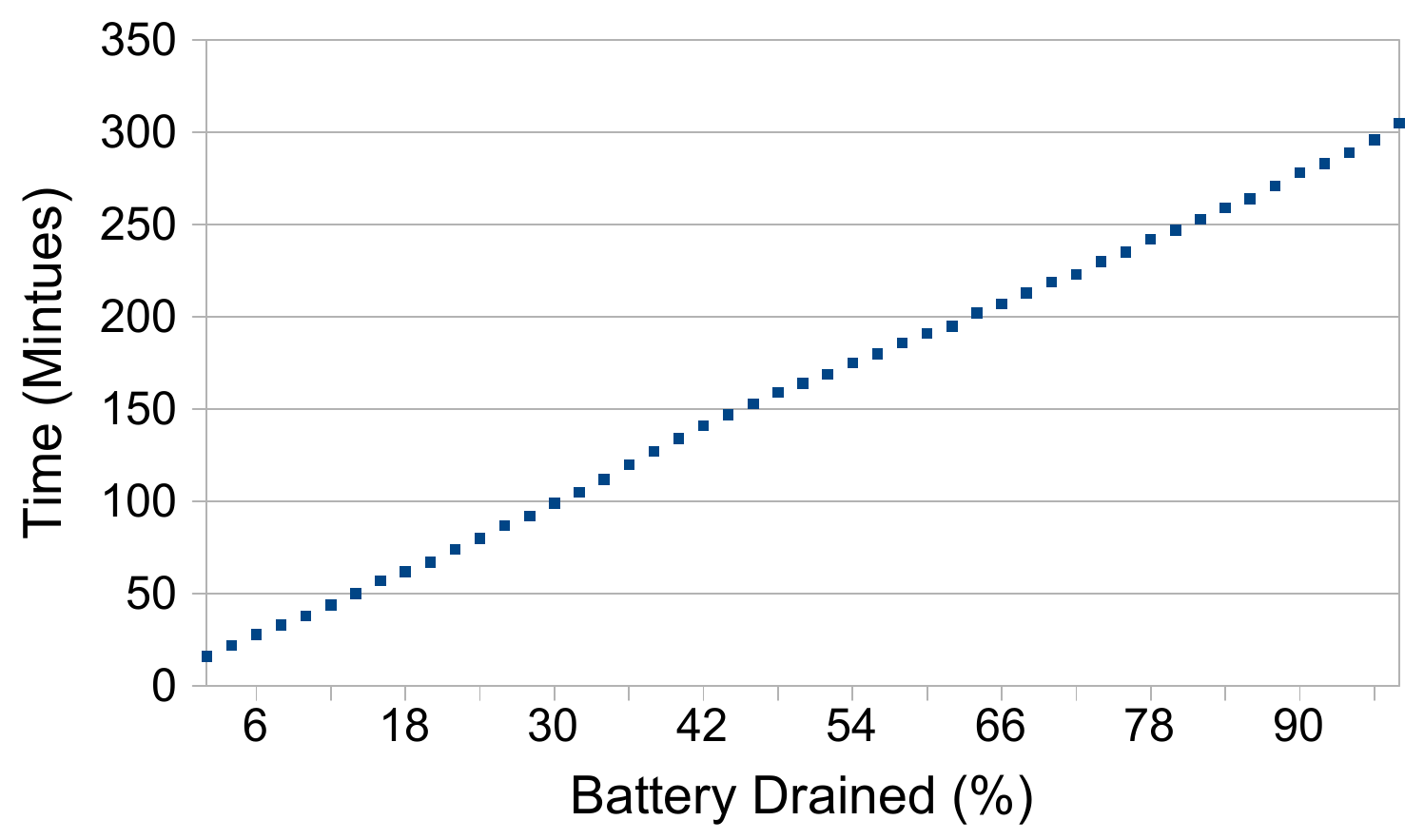}}
\end{minipage}

\\

\begin{minipage}[b]{0.33\textwidth}
 \subfigure[Energy drain of database data encryption, insertion and deletion
 attack over full battery life]{
\label{fig:dbencr}
\includegraphics[width=2.2in]{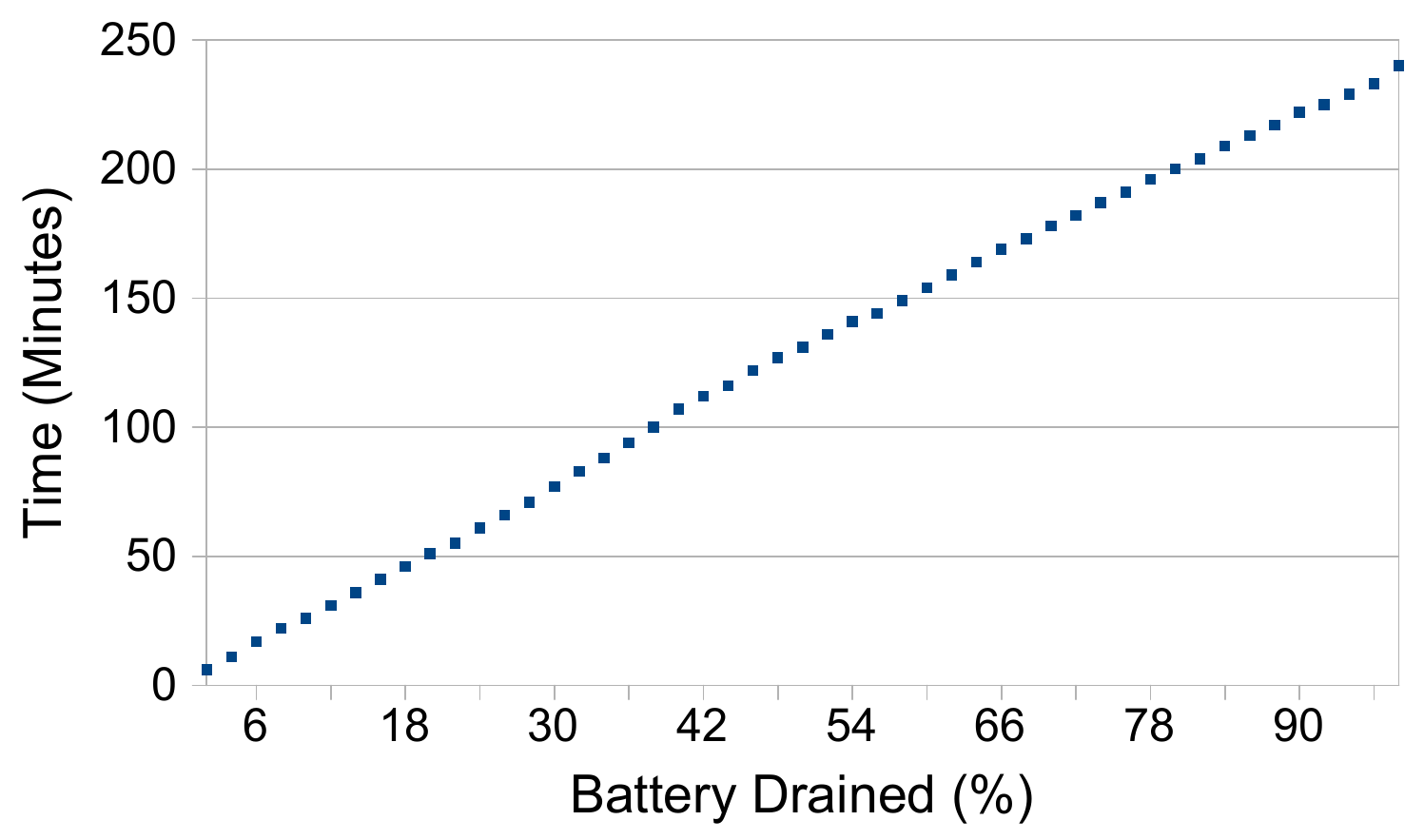}}
\end{minipage}
&
\begin{minipage}[b]{0.33\textwidth}
 \subfigure[Energy drain of web app attack over full battery life]{
\label{fig:web}
\includegraphics[width=2.2in]{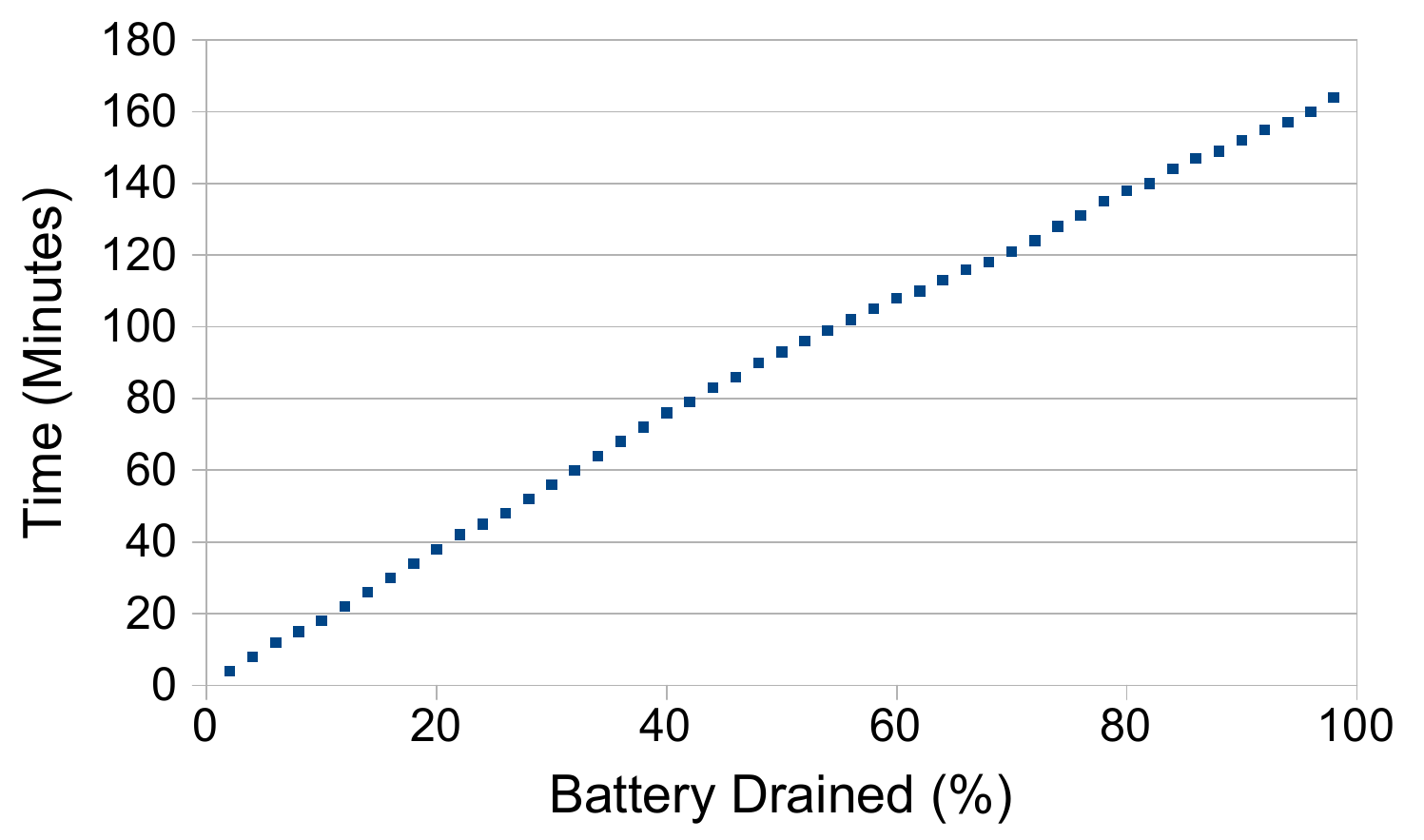}}
\end{minipage}
&
\begin{minipage}[b]{0.33\textwidth}
 \subfigure[Energy drain of most efficient attack over full battery life]{
\label{fig:finalattack}
\includegraphics[width=2.2in]{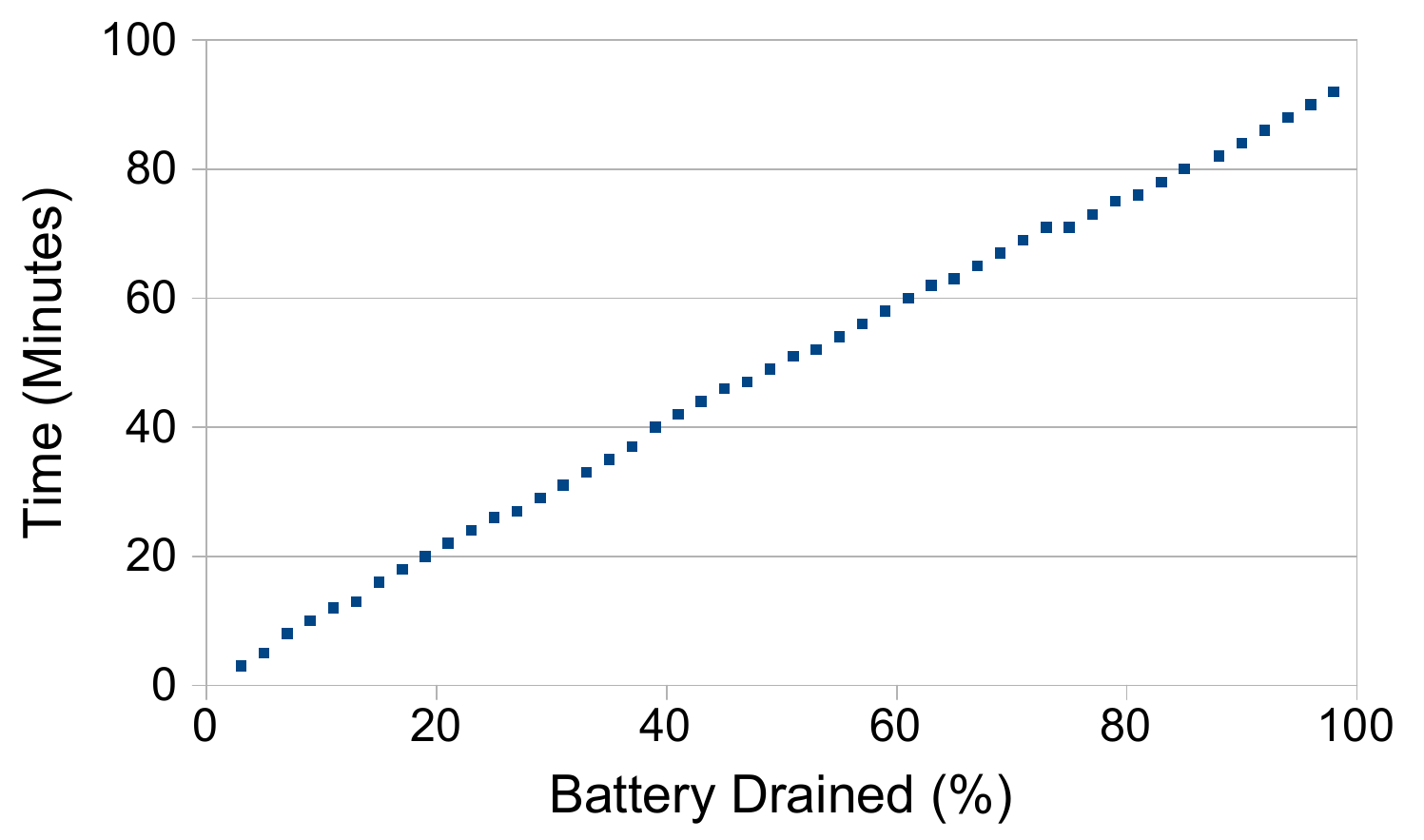}}
\end{minipage}

\end{tabular}
\label{fig:taskprofile} \caption{Exploitation using software elements.}
\end{figure*}

%% file: related.tex

\paragraph{Malicious Battery Draining Apps.} Even though, we are not aware of any malicious implementation of energy attacks
for smartphones running iOS and Android, we have found apps that may carry out
energy attacks inadvertently or as a side effect of other malicious activities
such as data harvesting or a side effect of suboptimized code. Some of the apps
we have found to be draining energy of a device by operations that are not
needed:

\begin{itemize}
  \item ``Flashlight''~\cite{flashlight} by Quick Switch on iPhone 4S: It acquires GPS, sends the
  location to the website perhaps for  location-specific ads and does
  not release it for sometime. It has been found to drain the battery of the
  phone overnight, if the app is not stopped.
  \item Waze~\cite{waze} app does not release the ``GPS'' for several minutes even though
  the user has stopped using the app. That leads to battery drainage.
\end{itemize}

\paragraph{Energy Analysis on Mobile Devices} Over the past decade, the battery usage analysis of mobile devices has attracted a lot of attentions. Neugebauer and McAuley~\cite{Neugebauer:2001} had suggested that using performance counter data to accurately account power consumption for laptops and mobile devices. To detect energy-greedy mobile malware such as WiFi faker, Kim et al.~\cite{Kim:2008} proposed a power-aware malware detection framework by collecting application power consumption signatures. Recently, Carroll and Heiser~\cite{Carroll:2010:APC} has systematically measured the energy usage and battery lifetime of a modern mobile device, at the major subsystem level such as graphics, GSM, WIFI, with a wide range of usage scenarios.  EPROF~\cite{pathak2012energy} is an energy profiler that can quatitively measure the battery usage of smartphone apps, from system call tracing perspective based on their earlier research~\cite{Pathak:2011:FPM}. Similarly, AppScope~\cite{Yoon:2012}, an Android-based energy metering system, uses an event-driven kernel activity monitoring to meter the application energy usage. For troubleshooting battery drain issues, eDoctor~\cite{Ma:2013} is a tool that can identify an abnormal app and suggest appropriate repair solution to users.

Compared to all these works, the substantial difference is we have different research goals. In particular, Kim et al.~\cite{Kim:2008} focused on malware detection, Carroll and Heise~\cite{Carroll:2010:APC} focused on identifying the critical components that consumes the most energy and then giving directions to power management, EPROF~\cite{pathak2012energy} and AppScope~\cite{Yoon:2012} focused on the profiling of energy usage of mobile apps, eDoctor~\cite{Ma:2013} focused on the energy diagnosis, whereas we focus on identify the stealthy and promising approach to quickly drain the battery from offensive perspective. On the other hand, all these techniques can facilitate us for the better measurement of the energy usage for each specific attack.

\paragraph{Energy Analysis on Servers} While energy attack has not been fully explored in the mobile devices, there are works that focus on the server side. Specifically, Wu et al.~\cite{wu2011energy} studied the security aspect of modern computer power management, and designed an energy attack that exploits a standalone server system, in a similar fashion to the DoS attack. Not for offensive purposes, Li et al.~\cite{Li:2012} studied the power consumption on modern enterprise storage systems for a better power efficient design. In addition, for energy savings purposes, Sueur and Heiser~\cite{LeSueur:2011} analyzed the modern energy saving techniques with various workloads and shed lights on how we should analyze the usage of power-management mechanisms.

%% file: conclusion.tex

Energy is a critical resource in mobile devices, and thus the emerging
dependence on smartphones for critical tasks and activities is also tied to
energy availability on the device. In this paper, we systematically analyze the
attack vectors from hardware, software, and network communication perspective to
drain the battery of a smartphone. In this paper, we have demonstrated  that
energy attack on smartphones is practical, and there are many incentives to
carry out these attacks. We have studied the attack space, the taxonomy and the  elements needed
to build exploits for such attacks. We have also designed a number of concrete
exploits, and our experimental results show how individual components as well as
their combinations may be used to drain off the battery and mobile device users
should be aware of these attacks.

%% file: paper.bbl
\begin{thebibliography}{10}

\ignore{
\bibitem{QEMU}
{QEMU: an open source processor emulator}.
\newblock {\em http://www.qemu.org/}.
}



\bibitem{And04analyzingmemory}
G.~Balakrishnan and T.~Reps.
\newblock Analyzing memory accesses in x86 executables.
\newblock In {\em Proceedings of International Conference on Compiler
  Construction (CC'04)}, pages 5--23. Springer-Verlag, 2004.

\bibitem{DIVINE}
G.~Balakrishnan and T.~Reps.
\newblock Divine: Discovering variables in executables.
\newblock In {\em Proc. of VMCAI 2007}, Nice, France, 2007. ACM
  Press.


\bibitem{Berdine07}
J.~Berdine, C.~Calcagno, B.~Cook, D.~Distefano, P.~W. O'Hearn, T.~Wies, and
  H.~Yang.
\newblock Shape analysis for composite data structures.
\newblock In {\em CAV 2007}.

\bibitem{dec094}
P.~T. Breuer and J.~P. Bowen.
\newblock Decompilation: The enumeration of types and grammars.
\newblock {\em ACM Trans. Program. Lang. Syst.}, 16(5):1613--1647, 1994.

\bibitem{brumley:2006:alias}
D.~Brumley and J.~Newsome.
\newblock Alias analysis for assembly.
\newblock Technical Report CMU-CS-06-180, Carnegie Mellon University School of
  Computer Science, 2006.

\bibitem{point-to-lcpc94}
M.~G. Burke, P.~R. Carini, J.-D. Choi, and M.~Hind.
\newblock Flow-insensitive interprocedural alias analysis in the presence of
  pointers.
\newblock In {\em Proc. of the 7th International Workshop on Languages
  and Compilers for Parallel Computing}, pages 234--250, London, UK, 1995.
  Springer-Verlag.

\bibitem{Juan:ndss10}
J.~Caballero, N.~M. Johnson, S.~McCamant, and D.~Song.
\newblock Binary code extraction and interface identification for security
  applications.
\newblock In {\em Proceedings of the 17th Annual Network and Distributed System
  Security Symposium (NDSS'10)}, San Diego, CA, February 2010.

\bibitem{dawn:polyglot:CCS207}
J.~Caballero and D.~Song.
\newblock Polyglot: Automatic extraction of protocol format using dynamic
  binary analysis.
\newblock In {\em Proceedings of the 14th {ACM} Conference on Computer and and
  Communications Security ({CCS'07})}, pages 317--329, Alexandria, Virginia,
  USA, 2007.


\bibitem{osdi08-klee}
C.~Cadar, D.~Dunbar, and D.~Engler.
\newblock Klee: Unassisted and automatic generation of high-coverage tests for
  complex systems programs.
\newblock In {\em OSDI 2008}, San Diego, CA, 2008.

\bibitem{EXE:CCS'06}
C.~Cadar, V.~Ganesh, P.~M. Pawlowski, D.~L. Dill, and D.~R. Engler.
\newblock Exe: Automatically generating inputs of death.
\newblock In {\em Proc. of CCS 2006}, pages 322--335, Alexandria, Virginia, USA,
  2006. ACM.

\bibitem{shape-analysis-90}
D.~R. Chase, M.~Wegman, and F.~K. Zadeck.
\newblock Analysis of pointers and structures.
\newblock In {\em Proc. of PLDI 1990}, pages 296--310, White Plains,
  New York, USA, 1990. ACM.

\bibitem{revnic:eurosys10}
V.~Chipounov and G.~Candea.
\newblock Reverse engineering of binary device drivers with revnic.
\newblock In {\em Proc. of EuroSys 2010}, pages 167--180, Paris, France, 2010. ACM.

\bibitem{s2e}
V.~Chipounov, V.~Kuznetsov, and G.~Candea.
\newblock S2e: a platform for in-vivo multi-path analysis of software systems.
\newblock In {\em Proc. of the sixteenth international conference on
  Architectural support for programming languages and operating systems},
  ASPLOS XVI, pages 265--278, Newport Beach, California, USA, 2011. ACM.

\bibitem{chow:datalifetime:Security'04}
J.~Chow, B.~Pfaff, K.~Christopher, and M.~Rosenblum.
\newblock Understanding data lifetime via whole-system simulation.
\newblock In {\em Proc. of {USENIX} Security 2004}.

\bibitem{decompile}
C.~Cifuentes.
\newblock {Reverse Compilation Techniques}.
\newblock {\em PhD thesis, Queensland University of Technology}, 1994.

\bibitem{algorithms}
T.~Cormen, C.~Leiserson, R.~Rivest and C.~Stein
\newblock {\em Introduction to Algorithms (Third Edition)}, pages 603--610.
\newblock {\em The MIT Press}, London, England, 2009.

\bibitem{laika:osdi08}
A.~Cozzie, F.~Stratton, H.~Xue, and S.~T. King.
\newblock Digging for data structures.
\newblock In {\em Proc. of OSDI 2008}, pages 231--244, San Diego, CA, December, 2008.

\bibitem{Minos}
J.~R. Crandall, S.~F. Wu, and F.~T. Chong.
\newblock Minos: Architectural support for protecting control data.
\newblock {\em ACM Trans. Archit. Code Optim.}, 3(4):359--389, 2006.

\bibitem{cdecl}
J.~de~Boyne~Pollard.
\newblock The gen on function calling conventions.
\newblock http://homepage.ntlworld.com./jonathan.deboynepollard
  /FGA/function-calling-conventions.html.

\bibitem{debray98alias}
S.~K. Debray, R.~Muth, and M.~Weippert.
\newblock Alias analysis of executable code.
\newblock In {\em Proc. of POPL 1998},
  pages 12--24, 1998.

\bibitem{ret-into-libc-97}
S.~Designer.
\newblock ``return-to-libc" attack.
\newblock {\em Bugtraq}, August 1997.

\bibitem{shape-analysis-94}
A.~Deutsch.
\newblock Interprocedural may-alias analysis for pointers: beyond k-limiting.
\newblock In {\em Proc. of PLDI 1994}, pages 230--241, Orlando,
  Florida, USA, 1994. ACM.

\bibitem{virtuoso}
B.~Dolan-Gavitt, T.~Leek, M.~Zhivich, J.~Giffin, and W.~Lee.
\newblock Virtuoso: Narrowing the semantic gap in virtual machine
  introspection.
\newblock In {\em Proceedings of 2011 IEEE Symposium on Security and Privacy},
  pages 297--312, Oakland, CA, USA, 2011.

\bibitem{krugal:usenix07}
M.~Egele, C.~Kruegel, E.~Kirda, H.~Yin, , and D.~Song.
\newblock Dynamic spyware analysis.
\newblock In {\em Proc. of Usenix 2007}, June 2007.

\bibitem{Boomerang}
M.~V. Emmerik and T.~Waddington.
\newblock Using a decompiler for real-world source recovery.
\newblock In {\em Proc. of the 11th Working Conference on Reverse
  Engineering}, pages 27--36, 2004.

\bibitem{fuzz2}
J.~E. Forrester and B.~P. Miller.
\newblock An empirical study of the robustness of {Windows NT} applications
  using random testing.
\newblock In {\em Proc. of the 4th Conference on USENIX Windows Systems
  Symposium}, pages 6--6, Seattle, Washington, 2000. USENIX Association.


\bibitem{VMST}
Y.~Fu and Z.~Lin.
\newblock Space traveling across vm: Automatically bridging the semantic-gap in
  virtual machine introspection via online kernel data redirection.
\newblock In {\em Proceedings of the 2012 IEEE Symposium on Security and
  Privacy}, San Francisco, CA, May 2012.


\bibitem{Livewire:Garfinkel2003}
T.~Garfinkel and M.~Rosenblum.
\newblock {A virtual machine introspection based architecture for intrusion
  detection}.
\newblock In {\em Proc. of NDSS 2003}, February 2003.

\bibitem{grammar:fuzzy:pldi08}
P.~Godefroid, A.~Kiezun, and M.~Y. Levin.
\newblock Grammar-based whitebox fuzzing.
\newblock In {\em Proc. of PLDI 2008}, pages 206--215, Tucson, AZ,
  USA, 2008. ACM.

\bibitem{WhiteBoxFuzzy:2008}
P.~Godefroid, M.~Levin, and D.~Molnar.
\newblock Automated whitebox fuzz testing.
\newblock In {\em Proc. of NDSS 2008}, San Diego, CA, February 2008.

\bibitem{Guo:issta06}
P.~J. Guo, J.~H. Perkins, S.~McCamant, and M.~D. Ernst.
\newblock Dynamic inference of abstract types.
\newblock In {\em Proc. of ISSTA 2006}, pages 255--265, Portland, Maine, USA, 2006.
  ACM.


\bibitem{differential-slicing}
N.~M. Johnson, J.~Caballero, K.~Z. Chen, S.~McCamant, P.~Poosankam, D.~Reynaud,
  and D.~Song.
\newblock Differential slicing: Identifying causal execution differences for
  security applications.
\newblock In {\em Proceedings of the 2011 IEEE Symposium on Security and
  Privacy}, SP '11, pages 347--362. IEEE Computer Society, 2011.



\bibitem{gadget:2010}
C.~Kolbitsch, T.~Holz, C.~Kruegel, and E.~Kirda.
\newblock Inspector gadget: Automated extraction of proprietary gadgets from
  malware binaries.
\newblock In {\em Proc. of S\&P 2010}, Oakland, CA,
  May 2010.

\bibitem{TIE:NDSS11}
J.~Lee, T.~Avgerinos, and D.~Brumley.
\newblock Tie: Principled reverse engineering of types in binary programs.
\newblock In {\em Proc. of NDSS 2011}, San Diego, CA, February 2011.

\bibitem{point-to-fse99}
D.~Liang and M.~J. Harrold.
\newblock Efficient points-to analysis for whole-program analysis.
\newblock In {\em Proc. of the 7th European Software Engineering
  Conference held jointly with the 7th ACM SIGSOFT International Symposium on
  Foundations of Software Engineering (ESEC/FSE-7)}, pages 199--215, Toulouse,
  France, 1999. Springer-Verlag.

\bibitem{Rewards:NDSS10}
Z.~Lin, X.~Zhang, and D.~Xu.
\newblock Automatic reverse engineering of data structures from binary
  execution.
\newblock In {\em Proc. of NDSS 2010}, San Diego, CA, February 2010.

\bibitem{zlin:DSN10}
Z.~Lin, X.~Zhang, and D.~Xu.
\newblock Reuse-oriented camouflaging trojan: Vulnerability detection and
  attack construction.
\newblock In {\em Proc. of DSN-DCCS 2010}, Chicago, IL,
  USA, June 2010.

\bibitem{PIN}
{\sc C.-K. Luk, R. Cohn, R. Muth, H. Patil, A. Klauser, G. Lowney,
  S. Wallace, V.~J. Reddi, and K. Hazelwood}
\newblock Pin: Building customized program analysis tools with dynamic
  instrumentation.
\newblock In {\em Proc. of PLDI 2005}.

\bibitem{Manevich05}
R.~Manevich, E.~Yahav, G.~Ramalingam, and M.~Sagiv.
\newblock Predicate abstraction and canonical abstraction for singly-linked
  lists.
\newblock In {\em VMCAI}, Jan. 2005.

\bibitem{regionsismm}
M.~Marron, D.~Kapur, and M.~Hermenegildo.
\newblock Identification of logically related heap regions.
\newblock In {\em ISMM}, June 2009.

\bibitem{fuzz1}
B.~P. Miller, L.~Fredriksen, and B.~So.
\newblock An empirical study of the reliability of {UNIX} utilities.
\newblock In {\em Proc. of the Workshop of Parallel and Distributed
  Debugging}, pages 9--19, Academic Medicine, 1990.

\bibitem{Mitchell06}
N.~Mitchell.
\newblock The runtime structure of object ownership.
\newblock In {\em ECOOP}, July 2006.

\bibitem{TaintCheck}
J.~Newsome and D.~Song.
\newblock Dynamic taint analysis for automatic detection, analysis, and
  signature generation of exploits on commodity software.
\newblock In {\em Proc. of NDSS 2005}, San Diego, CA, February 2005.

\bibitem{point-to-07}
D.~J. Pearce, P.~H. Kelly, and C.~Hankin.
\newblock Efficient field-sensitive pointer analysis of c.
\newblock {\em ACM Trans. Program. Lang. Syst.}, 30(1):4, 2007.



\bibitem{POPL99}
G.~Ramalingam, J.~Field, and F.~Tip.
\newblock Aggregate structure identification and its application to program
  analysis.
\newblock In {\em Proceedings of the 26th ACM SIGPLAN-SIGACT Symposium on
  Principles of programming languages (POPL'99)}, pages 119--132, San Antonio,
  Texas, 1999. ACM.



\bibitem{cc:RepsB08}
T.~W. Reps and G.~Balakrishnan.
\newblock Improved memory-access analysis for x86 executables.
\newblock In {\em Proceedings of International Conference on Compiler
  Construction (CC'08)}, pages 16--35, 2008.


\bibitem{shape-analysis-repes99}
M.~Sagiv, T.~Reps, and R.~Wilhelm.
\newblock Parametric shape analysis via 3-valued logic.
\newblock In {\em Proc. of POPL 1999}, pages 105--118, San Antonio,
  Texas, USA, 1999. ACM.


\bibitem{Hovav:CCS'07}
H.~Shacham.
\newblock The geometry of innocent flesh on the bone: return-into-libc without
  function calls (on the x86).
\newblock In {\em Proceedings of the 14th ACM Conference on Computer and
  Communications Security}, pages 552--561, Alexandria, Virginia, USA, 2007.
  ACM.



\bibitem{Howard}
A.~Slowinska, T.~Stancescu, and H.~Bos.
\newblock Howard: A dynamic excavator for reverse engineering data structures.
\newblock In {\em Proceedings of the 18th Annual Network and Distributed System
  Security Symposium (NDSS'11)}, San Diego, CA, February 2011.

\bibitem{dual-slicing}
D.~Weeratunge, X.~Zhang, W.~N. Sumner, and S.~Jagannathan.
\newblock Analyzing concurrency bugs using dual slicing.
\newblock In {\em Proceedings of the 19th international symposium on Software
  testing and analysis}, ISSTA '10, pages 253--264, Trento, Italy, 2010. ACM.

\end{thebibliography}


\begin{thebibliography}{10}

\bibitem{waze}
Waze social gps maps and traffic.
\newblock https://play.google.com/store/apps/details?id=com.waze.

\bibitem{Carroll:2010:APC}
A.~Carroll and G.~Heiser.
\newblock An analysis of power consumption in a smartphone.
\newblock In {\em Proceedings of the 2010 USENIX conference on USENIX annual
  technical conference}, USENIXATC'10, pages 21--21, Berkeley, CA, USA, 2010.
  USENIX Association.

\bibitem{flashlight}
D.~Gilbert.
\newblock First ios malware discovered in apple's app store, 2012.
\newblock
  http://apple.slashdot.org/story/12/07/05/1727215/first-ios-malware-discovered-in-apples-app-store.

\bibitem{Kim:2008}
H.~Kim, J.~Smith, and K.~G. Shin.
\newblock Detecting energy-greedy anomalies and mobile malware variants.
\newblock In {\em Proceedings of the 6th international conference on Mobile
  systems, applications, and services}, MobiSys '08, pages 239--252, New York,
  NY, USA, 2008. ACM.

\bibitem{LeSueur:2011}
E.~Le~Sueur and G.~Heiser.
\newblock Slow down or sleep, that is the question.
\newblock In {\em Proceedings of the 2011 USENIX conference on USENIX annual
  technical conference}, USENIXATC'11, pages 16--16, Berkeley, CA, USA, 2011.
  USENIX Association.

\bibitem{Li:2012}
Z.~Li, K.~M. Greenan, A.~W. Leung, and E.~Zadok.
\newblock Power consumption in enterprise-scale backup storage systems.
\newblock In {\em Proceedings of the 10th USENIX conference on File and Storage
  Technologies}, FAST'12, pages 6--6, Berkeley, CA, USA, 2012. USENIX
  Association.

\bibitem{Ma:2013}
X.~Ma, P.~Huang, X.~Jin, P.~Wang, S.~Park, D.~Shen, Y.~Zhou, L.~K. Saul, and
  G.~M. Voelker.
\newblock edoctor: automatically diagnosing abnormal battery drain issues on
  smartphones.
\newblock In {\em Proceedings of the 10th USENIX conference on Networked
  Systems Design and Implementation}, nsdi'13, pages 57--70, Berkeley, CA, USA,
  2013. USENIX Association.

\bibitem{imessage}
D.~Murphy.
\newblock Industrious users performing imessage denial-of-service attacks,
  2013.
\newblock http://www.pcmag.com/article2/0,2817,2417267,00.asp.

\bibitem{Neugebauer:2001}
R.~Neugebauer and D.~McAuley.
\newblock Energy is just another resource: Energy accounting and energy pricing
  in the nemesis os.
\newblock In {\em Proceedings of the Eighth Workshop on Hot Topics in Operating
  Systems}, HOTOS '01, pages 67--, Washington, DC, USA, 2001. IEEE Computer
  Society.

\bibitem{pathak2012energy}
A.~Pathak, Y.~C. Hu, and M.~Zhang.
\newblock Where is the energy spent inside my app?: fine grained energy
  accounting on smartphones with eprof.
\newblock In {\em Proceedings of the 7th ACM european conference on Computer
  Systems}, pages 29--42. ACM, 2012.

\bibitem{Pathak:2011:FPM}
A.~Pathak, Y.~C. Hu, M.~Zhang, P.~Bahl, and Y.-M. Wang.
\newblock Fine-grained power modeling for smartphones using system call
  tracing.
\newblock In {\em Proceedings of the sixth conference on Computer systems},
  EuroSys '11, pages 153--168, New York, NY, USA, 2011. ACM.

\bibitem{Paul2010androidenergy}
K.~Paul and T.~Kundu.
\newblock Android on mobile devices: An energy perspective.
\newblock In {\em Computer and Information Technology (CIT), 2010 IEEE 10th
  International Conference on}, pages 2421--2426, 2010.

\bibitem{wu2011energy}
Z.~Wu, M.~Xie, and H.~Wang.
\newblock Energy attack on server systems.
\newblock In {\em Proceedings of the 5th USENIX conference on Offensive
  technologies}. USENIX Association, 2011.

\bibitem{Yoon:2012}
C.~Yoon, D.~Kim, W.~Jung, C.~Kang, and H.~Cha.
\newblock Appscope: application energy metering framework for android
  smartphones using kernel activity monitoring.
\newblock In {\em Proceedings of the 2012 USENIX conference on Annual Technical
  Conference}, USENIX ATC'12, pages 36--36, Berkeley, CA, USA, 2012. USENIX
  Association.

\bibitem{zhou2012dissecting}
Y.~Zhou and X.~Jiang.
\newblock Dissecting android malware: Characterization and evolution.
\newblock In {\em Security and Privacy (SP), 2012 IEEE Symposium on}, pages
  95--109. IEEE, 2012.

\end{thebibliography}
